\documentclass[12pt,final]{elsart}

  \newcommand{\preprint}{SACLAY--T06/148\\ECT*--06--19}

  \usepackage{latexsym,bm,amsmath,amssymb,amsfonts}
  \usepackage{epsfig,graphics,graphicx,subfigure}
  \usepackage{cite}
  \usepackage{slashed}

 \def\simge{\mathrel{%
   \rlap{\raise 0.511ex \hbox{$>$}}{\lower 0.511ex \hbox{$\sim$}}}}
\def\simle{\mathrel{
   \rlap{\raise 0.511ex \hbox{$<$}}{\lower 0.511ex \hbox{$\sim$}}}}

 \long\def\comment#1{ }
  \newcommand{\dif}{{\rm d}}
  \newcommand{\dY}{\dif Y}
  \newcommand{\abar}{\bar{\alpha}_s}
  
  \newcommand{\del}{\partial}
  \newcommand{\lan}{\langle}
  \newcommand{\ran}{\rangle}
  \newcommand{\mcal}{\mathcal}
  
  \newcommand{\rmi}{{\rm i}}
  \newcommand{\rmL}{{\rm L}}
  \newcommand{\rmR}{{\rm R}}

  \newcommand{\order}[1]{\mcal{O}{(#1)}}
  \newcommand{\nn}{\nonumber\\}
  \newcommand{\coll}[1]{\{#1\}}

  \newcommand{\beq}{\begin{eqnarray}}
  \newcommand{\eeq}{\end{eqnarray}}
  \newcommand{\avg}[1]{\left\langle #1 \right\rangle}

\begin{document}

\begin{flushright}
{\small \preprint}
\end{flushright}
\vspace{1.5cm}

\begin{frontmatter}\parbox[]{16.0cm}{\begin{center}
\title{\rm \LARGE One--dimensional model for
QCD at high energy}


\author{E.~Iancu $^{\rm a,1,2}$},
\author{J.T.~de Santana Amaral $^{\rm b,1}$},
\author{G.~Soyez $^{\rm a,1,3}$},
\author{D.N.~Triantafyllopoulos $^{\rm c,1}$}

\address{$^{\rm a}$ Service de Physique Th\'eorique de Saclay,
CEA/DSM/SPhT, F-91191 Gif-sur-Yvette, France}
\address{$^{\rm b}$ Instituto de F\'isica, Universidade Federal do Rio
Grande do Sul, 91501 Porto Alegre (RS), Brazil}
\address{$^{\rm c}$ ECT$^*$, Villa Tambosi, Strada delle Tabarelle
286, I-38050 Villazzano (TN), Italy}

\thanks{{\it E-mail addresses:}
iancu@dsm-mail.cea.fr (E.~Iancu), thiago.amaral@ufrgs.br (J.T.~de
Santana Amaral), gsoyez@dsm-mail.cea.fr (G.~Soyez), dionysis@ect.it
(D.N.~Triantafyllopoulos).}
\thanks{Membre du Centre National de la Recherche Scientifique
(CNRS), France.}
\thanks{On leave from the Fundamental Theoretical Physics group
of the University of Li\`ege.}

\begin{abstract}
We propose a stochastic particle model in (1+1)--dimensions, with
one dimension corresponding to rapidity and the other one to the
transverse size of a dipole in QCD, which mimics high--energy
evolution and scattering in QCD in the presence of both saturation
and particle--number fluctuations, and hence of Pomeron loops. The
model evolves via non--linear particle splitting, with a non--local
splitting rate which is constrained by boost--invariance and
multiple scattering. The splitting rate saturates at high density,
so like the gluon emission rate in the JIMWLK evolution. In the mean
field approximation obtained by ignoring fluctuations, the model
exhibits the hallmarks of the BK equation, namely a BFKL--like
evolution at low density, the formation of a traveling wave, and
geometric scaling. In the full evolution including fluctuations, the
geometric scaling is washed out at high energy and replaced by
diffusive scaling. It is likely that the model belongs to the
universality class of the reaction--diffusion process. The analysis
of the model sheds new light on the Pomeron loops equations in QCD
and their possible improvements.
\end{abstract}
\end{center}}

\end{frontmatter}

\newpage

\section{Introduction}\label{sect-intro}
\setcounter{equation}{0}

Much of the recent progress towards understanding the dynamics of
QCD at high energies comes from the observation \cite{IMM04,IT04}
that the QCD evolution (at least, in its `leading--logarithmic
approximation' with respect to the energy logarithm $\ln s$) is a
relatively simple classical stochastic process in the universality
class of the `reaction--diffusion process'.

In its canonical formulation (see, e.g., Refs.
\cite{Gardiner,Saar}), the reaction--diffusion process $A
\rightleftharpoons AA$ involves particles of type $A$ located at the
sites of an infinite, one--dimensional, lattice, which are endowed
with the following dynamics: a particle can locally split into two
($A\to AA$), two particles can recombine into one ($AA\to A$), and a
particle can diffuse from one site to the adjacent sites. But the
associated {\em universality class} covers a wide variety of
stochastic processes, whose detailed microscopic dynamics can be
different from those of the `canonical' realization described above,
but which share the same basic ingredients, leading to an {\em
effective} one--dimensional dynamics with local growth and diffusion
in the dilute regime, saturation of the occupation numbers at high
density, and fluctuations in the particle number. These ingredients
appear to be sufficient to ensure a {\em universal} behaviour for
all such processes in specific limits (large time, weak noise), as
empirically demonstrated by the experience with a large number of
models \cite{Saar}, and conceptually understood from general
arguments \cite{BD97,BDMM}.

In particular, in the context of high--energy QCD, the `particles'
correspond to gluons (or dipoles), the `time' axis corresponds to
the rapidity $Y\sim \ln s$, and the one--dimensional `spatial' axis
is the logarithm of the gluon transverse momentum (or the dipole
transverse size). Furthermore, the `particle splitting' corresponds
to the BFKL evolution \cite{BFKL,AM94}, the `recombination', to the
non--linear effects responsible for gluon saturation
\cite{GLR,MQ85,MV,B,K,JKLW,CGC,W}, and the `diffusion', to the
non--locality of the BFKL kernel and the various gluon, or dipole,
vertices in the transverse space. Of course, the actual QCD problem
is much more complicated than just particle splitting and merging:
it involves color degrees of freedom, whose role appears to be
irreducible at high density; also, gluon saturation (as described by
the JIMWLK equation \cite{JKLW,CGC,W}) does not occur via
`mechanical' recombination, but rather as a consequence of coherent,
strong field, effects \cite{SAT}; besides, the transverse plane has
two dimensions and, moreover, the theoretical descriptions typically
involve two types of transverse coordinates
--- the transverse size (or momentum) and the position in impact
parameter space ---, so the relevant transverse phase--space is
truly four--dimensional. In spite of this, one can argue that, under
suitable approximations, the dominant high--energy dynamics at fixed
impact parameter is indeed one--dimensional, with the relevant
dimension being the transverse size (or momentum), as alluded to
above \cite{IMM04,IT04}.

The universality of the reaction--diffusion process is especially
interesting in this QCD context, since in spite of considerable work
and progress over the recent years
\cite{LipatovS,B,K,JKLW,CGC,W,SAT,MS04,IMM04,IT04,MSW05,IT05,KL05,KL3,BIIT05,Braun05,BREM,Balit05},
a complete theory for the high--energy evolution and scattering in
QCD is still lacking. Moreover, even the approximate versions of
this theory that are currently known, like the `Pomeron loop'
equations proposed in Refs. \cite{IT04,MSW05,IT05,BIIT05}, or the
`self--dual' \cite{KL3} effective Hamiltonian of Refs.
\cite{BREM,Balit05}, appear to be too complicated to deal with in
practice, and may also suffer from internal inconsistencies
\cite{IST05}. In view of this, it is particularly rewarding that
some fundamental aspects of QCD at high energy, like the asymptotic
behaviour of the scattering amplitudes (in particular, the breakdown
of `geometric' scaling \cite{geometric,SCALING,MT02,DT02,MP03} and
the emergence of a new, `diffusive', scaling at very high energies),
could have been inferred from the correspondence with the
reaction--diffusion process in statistical physics
\cite{IMM04,IT04,GS05,EGBM05,BDMM,HIMST06,MSX06}.

But this correspondence has its own, rather strong, limitations,
which could be overcome only through deeper investigations of the
actual, microscopic, dynamics. In view of the difficulty to perform
such studies directly in QCD, one nowadays assists at a regain of
interest in {\em stochastic particle models}
\cite{LL05,Stasto05,GS05,EGBM05,SX05,KozLev06,BIT06,BMMSX06}, which
aim at simulating the `Pomeron loops dynamics' --- by which we main
the high--energy evolution and scattering in the presence of both
saturation and particle--number fluctuations --- in lower dimensions
(zero or one transverse dimensions). Such models are interesting not
only because they allow for explicit solutions (even analytic ones,
in the case of zero dimensions \cite{SX05,KozLev06,BIT06,BMMSX06}),
and thus permit us to familiarize ourselves with the physical
consequences of the Pomeron loops, but also because they provide
explicit evolution equations, and thus give us some insight into the
structural aspects that one should expect in QCD. Besides, if
sufficiently rich, a model may also allow us to study {\em
non}--universal aspects, like the influence of the initial
conditions at low energy, or the details of the transition from
geometric to diffusive scaling.

However, in order to meet with these desiderata a model should
contain as much as possible of the actual QCD dynamics, to the
extent that this can be made consistent with the limitations
inherent in the model (like its low dimensionality and the lack of
color degrees of freedom). For instance, the model should respect
general symmetry properties, like the {\em boost--invariance} of the
evolution equations, which in fact is one of the main physical
arguments in favour of Pomeron loops \cite{IM031,MS04,KL3,BIIT05}.
(This condition is sometimes reformulated as the `self--duality' of
the evolution Hamiltonian \cite{KL3,BIIT05,BREM}.) Also, the model
should incorporate some general properties of QCD at high energy,
like the fact that a particle which scatters off a dense system
undergoes {\em multiple scattering}. Moreover, in the limiting cases
where the QCD equations are presently known, the model should
reproduce suitable versions of these equations: it should be similar
to the QCD dipole picture \cite{AM94,IM031} in the dilute regime,
and it should reduce to a suitable version of the BK equation, or of
the Balitsky--JIMWLK hierarchy\footnote{One should perhaps recall at
this point that the BK equation and the Balitsky hierarchy are
essentially equivalent at large--$N_c$, which is the limit of
interest for the particle models.} \cite{B,K,JKLW,CGC,W}, in the
high--density regime where the particle--number fluctuations become
unimportant.

Together, the above conditions turn out to be very constraining, and
it is not so easy to construct a model which fulfills all these
constraints. In fact, most of the models proposed so far are
directly inspired by the canonical reaction--diffusion process
\cite{LL05,Stasto05,GS05,EGBM05,SX05,KozLev06,BMMSX06} (which is
also closely related \cite{BMMSX06} to the old `Reggeon field
theory') and as such, not only they cannot be used to study
deviations from universality, but also they have difficulties to
accommodate multiple scattering: to be consistent with
boost--invariance, such models must assume that a projectile
particle can undergo only single scattering, however dense the
target is.

A noticeable exception in that sense is a zero--dimensional model
originally proposed by Mueller and Salam \cite{AMSalam95}, in an
attempt to include saturation effects in the dipole picture, which
has been recently reconsidered in Ref. \cite{BIT06} (see also Ref.
\cite{KL5}). In this model, boost--invariance and multiple
scattering (in the eikonal approximation) are used as building
principles, and it turns out that --- due to simplifications
associated with the lack of transverse dimensions --- these two
conditions are almost enough to uniquely fix the evolution law. The
only additional assumption, which is fully consistent with the known
situation in QCD, is that the evolution proceeds via particle
splitting {\em alone} (no recombination), with one new particle
being emitted per unit rapidity. Remarkably, the particle emission
rate which emerges from these constraints is {\em non--linear} in
the number of preexisting particles and {\em saturates} at some
maximal value when this number is large enough, as a consequence of
multiple scattering. These features are appealing since rather
similar to gluon saturation in the framework of the JIMWLK
evolution.

But the lack of transverse dimensions greatly restricts the spectrum
of physical problems that can be studied in the framework of the
model: the only non--trivial problem which can be addressed in zero
dimensions \cite{AMSalam95,BIT06} is that of the approach of the
$S$--matrix towards the black disk limit $S=0$. But the most
interesting problems in the context of the high--energy evolution
are those related to the coexistence of two regimes, one dense and
one dilute (at low, and respectively, high transverse momenta),
separated by the saturation momentum. In order to consider such
problems one needs at least one transverse dimension --- that
expressing the gluon momentum, or the dipole size. It is our purpose
in this paper to present an extension of the particle model in Refs.
\cite{AMSalam95,BIT06} to one transverse dimension, denoted as $x$,
to be eventually identified --- in the correspondence with QCD
--- to the logarithm of the inverse size of a
dipole: $x \leftrightarrow \ln(r_0^2/r^2)$, where $r$ is the generic
dipole size and $r_0$ is some arbitrary scale of reference
(typically, the size of the dipole which started the evolution at
$Y=0$).

The model that we shall consider is constrained by the same physical
assumptions as in Refs. \cite{AMSalam95,BIT06} ---
boost--invariance, multiple scattering and evolution via particle
splitting ---, but in the presence of the transverse dimension these
assumptions are not sufficient anymore to completely fix the
dynamics. Yet, as we shall see, they strongly constrain this
dynamics: the only non--trivial (class of) solution(s) to these
constraints that we shall find is the one which appears as the
natural, although not necessarily obvious, generalization of the
original model in zero dimensions. Specifically, the elementary
particle--particle scattering amplitude is now {\em non--local} in
$x$, although very {\em short--ranged}, corresponding to the fact
that QCD favors the interaction between dipoles with similar sizes.
Accordingly, the rate for particle emission at a particular site $x$
(i.e., for emitting dipoles of a given size $r$) is found to be
proportional to the scattering amplitude between the new particle
and the preexisting ones, and thus it depends upon the occupation
numbers $n(y)$ at all the other sites $y$. Once again, this emission
rate saturates at its maximal value when the occupation numbers at
the site of interest, or in its vicinity, become large enough.

The one--dimensional model endowed with this dynamics develops a
very interesting structure, which is formally similar to that of the
corresponding model in zero dimensions \cite{BIT06}, but whose
physical consequences are considerably richer, and also much closer
to QCD. The evolution Hamiltonian\footnote{In the context of a
stochastic particle model, a Hamiltonian can be identified by
rewriting the master equation in operator form; see Sect.
\ref{sect-const} below for details.}, which is self--dual (as it
should, given the built-in boost--invariance), can be recognized as
an extension of the JIMWLK Hamiltonian which allows for
particle--number fluctuations. The equations for the scattering
amplitudes generated by this Hamiltonian provide an intuitive
generalization of the Balitsky--JIMWLK hierarchy in which the
projectile and the target are symmetrically treated. The general
structure of these equations is in fact similar to that of the
`Pomeron loop equations' proposed in QCD at large $N_c$
\cite{IT04,MSW05,IT05}, in the sense of including three types of
terms: linear terms responsible for the BFKL evolution, source terms
describing particle--number fluctuations in the target (or,
equivalently, saturation effects in the projectile), and non--linear
terms corresponding to saturation effects in the target, which
ensure the unitarization of the scattering amplitudes at high
energy.

At the same time, some interesting differences are found between the
structure of the fluctuation terms in the toy--model equations and,
respectively, the `Pomeron loop' equations in QCD, which invites us
to a more careful comparison between these two sets of equations.
This study (to be detailed in the Appendix) shows that the observed
differences reflect different ways to organize perturbation theory
in the weak scattering regime, which are essentially equivalent, in
the sense of providing the same dominant behaviour at high energy.
This being said, the toy--model equations appear to be more
symmetric in their treatment of the target versus projectile, and it
is likely that a similar symmetry should hold in the complete
equations of QCD, for which the toy model provides some inspiration.

Furthermore, the presence of a transverse dynamics in the toy model
allows for explicit comparisons with the corresponding dynamics in
QCD and, more generally, for studies of the universality. It appears
that the linearized version of the toy--model equations (as
appropriate in the dilute regime) is very close to the BFKL
dynamics, whereas the mean--field version of the same equations (in
which the expectation values are assumed to factorize) is similarly
close to the BK dynamics. Namely, in the dilute regime, the solution
shows the characteristic BFKL pattern \cite{BFKL} (exponential
increase in $Y$, exponential decrease in $x$ with some `anomalous
dimension', diffusion in $x$ with diffusive radius
$\propto\sqrt{Y}$), and even approaches `color transparency' at very
large values of $x$. After including non--linear terms in the
mean--field approximation, one observes saturation, the formation of
a traveling wave, and geometric scaling
--- all that being indeed very similar to QCD \cite{SCALING,MT02,MP03}.

A stochastic particle model with the linear and mean--field
behaviour aforementioned is expected to belong to the universality
class of the reaction--diffusion process, and this will be indeed
confirmed by a numerical study of our model: the position of the
traveling fronts (the toy--model analog of the logarithm of the
saturation momentum) is thus found to be a random quantity with a
dispersion increasing linearly with $Y$, and the shape of the
average amplitude is seen to approach diffusive scaling at very
large $Y$. Our present numerical analysis is only exploratory --- it
is merely intended to verify the universal features of the dynamics
---, but we plan to perform a more detailed such a study in a
further publication, and thus also study some non--universal
aspects.

To summarize, although conceptually rich and relatively close to QCD
(at least, in the limiting cases in which the comparison is
possible), the model that we shall introduce here remains relatively
simple and allows for systematic studies, including numerical ones.
This model could therefore serve as a playground for testing new
ideas concerning either the formal structure of the high--energy
evolution equations in QCD, or the effects of the evolution with
Pomeron loops on various physical processes.

The paper is organized as follows: In Sect. \ref{sect-const} we
shall describe the construction of the model from the underlying
physical assumptions, present the master equation for the
probabilities, and discuss some of its general properties. In Sect.
\ref{sect-observ}, we shall use the master equation to deduce
evolution equations for physical observables, like the $k$--body
particle densities and the scattering amplitudes. Then, in Sect.
\ref{sect-analytics}, we shall present an analytic study of these
equations, which will successively focus on the linear
approximation, the mean--field approximation, and the qualitative
aspects of the full evolution with fluctuations. These analytic
considerations will be then illustrated by the numerical results in
Sect. \ref{sect-numerics}, which will also substantiate our
expectation that the model should belong to the universality class
of the reaction--diffusion process. Finally, Sect. \ref{sect-concl}
contains our conclusion and a brief account of the analysis in the
Appendix.

\section{Construction of the model}\label{sect-const}
\setcounter{equation}{0}

In this section, we shall formulate our stochastic particle model
for high--energy evolution and scattering in QCD. As explained in
the Introduction, this is a $(1+1)$--dimensional model, where one of
the dimensions refers to the rapidity $Y$ (the logarithm of the
energy), which plays the role of a `time' for the high--energy
evolution, while the other one is a spatial dimension --- the
position of the particle along an infinite one--dimensional axis.
For the analogy with QCD, one should keep in mind that this spatial
dimension corresponds to a {\em size} in QCD, more precisely to the
logarithm of the inverse of the transverse size of a dipole. For the
sake of the presentation, we shall first describe the discretized
version of the model, where the spatial dimension is assimilated to
a one--dimensional lattice, with lattice points labeled by $i$. This
is, of course, also the version of the model which is best suited
for the numerical implementation to be eventually discussed.
However, as we shall later demonstrate, the continuum limit of the
model is well--defined and rather straightforward to obtain.

A {\em system} of particles is then defined by specifying the number
of particles $n_i$ (or `occupation numbers') at all the sites of the
lattice (some of these sites could be empty). With increasing
rapidity, the composition of the system can randomly vary, via
particle splitting according to a law to be shortly specified.
Hence, the particle distribution is stochastic, and the state of the
system at rapidity $Y$ is described by the probability
$P(\coll{n},Y)$ to find a given configuration $\{n\}\equiv \{\dots,
n_i, n_{i+1},\dots\}$ of particles.

In order to mimic a scattering problem, we shall consider two types
of particles, say particles of type `right' ($R$) and particles of
type `left' ($L$), which coexist at the lattice sites and interact
with each other via generally non--local interactions (see below for
details). The particle labels `$R$' and `$L$' refer, of course, to
the fact that in an actual scattering problem, the two colliding
systems propagate in opposite directions along the collision
axis\footnote{One should not confound this collision axis with the
spatial dimension in the toy model, which rather corresponds to
transverse degrees of freedom in QCD, as explained before.}.

In what follows, we shall assume that the average $S$--matrix
element for the elastic scattering can be given the factorized form
  \beq\label{eq-stotal}
   \lan S \ran_Y =
   \sum_{\coll{n},\coll{m}}
   P_{\rmR}(\coll{n},Y-Y_0)\, P_{\rmL}(\coll{m},Y_0)\,
   {\mcal S}(\coll{n},\coll{m}),
  \eeq
where $Y$ is the total rapidity separation between the two systems
and is divided between the `target' (the right mover), which has a
rapidity $Y-Y_0$ in the chosen frame, and the `projectile' (the
left--mover), which has a rapidity $-Y_0$ (with $Y_0>0$).
Furthermore, $P_{\rmR}(\coll{n},Y-Y_0)$ and $P_{\rmL}(\coll{m},Y_0)$
are probability distributions describing the two systems at the time
of scattering, and ${\mcal S}(\coll{n},\coll{m})$ is the $S$--matrix
for the elastic scattering between two given configurations
$\coll{n}$ and $\coll{m}$ of the target and, respectively, the
projectile. Finally, the sum in Eq.~(\ref{eq-stotal}) runs over all
the possible configurations in both systems.

A factorization similar to Eq.~(\ref{eq-stotal}) is known to hold in
QCD for onium--onium scattering within the dipole picture
\cite{AM95,IM031}. More generally, such a factorization emerges from
a more fundamental, quantum, description in terms of light--cone
wavefunctions whenever the single--particle states are eigenstates
of the $S$--matrix operator (which is our implicit assumption here).

On physical grounds, the average $S$--matrix must be independent of
the rapidity divider $Y_0$, i.e., upon the choice of a Lorentz
frame, which implies (with short--handed notations):
 \beq\label{eq-LorInv}\hspace*{-.5cm}
 0\,=\,  \frac{\dif\avg{S}}{\dif Y_0}\,=
   \sum_{\coll{n},\coll{m}}\,
   \Big(   P_{\rmR}(Y-Y_0)\,
   \frac{\del P_{\rmL}(Y_0)}{\del Y_0}\,+\,
   \frac{\del P_{\rmR}(Y-Y_0)}{\del Y_0}\, P_{\rmL}(Y_0)\Big)
   {\mcal S}(\coll{n},\coll{m})\,.
  \eeq
This condition, which relates properties of the scattering to those
of the evolution with $Y$, represents a rather strong constraint on
the latter, and can be even used to fix the structure of the
evolution law under some additional assumptions. Let us enumerate
here the main assumptions that we shall rely on in that respect:

\vspace*{.2cm}
\begin{enumerate}
\item[\tt(i)]{\em Multiple scattering in the eikonal approximation.}
Let $\sigma_{ij}=1-\tau_{ij}$ denote the $S$--matrix describing the
scattering between two elementary particles with positions $i$ and
$j\,$; here, $\tau_{ij}$ is the corresponding $T$--matrix and is
assumed to be real, as expected for the dominant behaviour at high
energy. Each particle from the target can scatter off all the
particles in the projectile, and different particles from the target
scatter independently off each other. Then, the S--matrix for given
configurations of the target and the projectile reads:
 \beq\label{eq-sconf}
    {\mcal S}(\coll{n},\coll{m})\,=\,
   \prod_{i,j} \sigma_{ij}^{n_i m_j}.
  \eeq
\item[\tt(ii)]{\em One particle emission per evolution step.}
We shall assume that, when increasing rapidity in one step ($Y \to
Y+\dY$), the evolution consists in the emission of a single
additional particle, with a probability which generally depends upon
all the occupation numbers in the system. That is, the additional
particle is emitted {\em coherently} from all the previous ones.
\end{enumerate}
\vspace*{.2cm}

Both these assumptions are natural from the perspective of
perturbative QCD at high energy: The eikonal approximation is the
standard method to resum multiple scattering in QCD at high--energy,
for both gluons (see, e.g., Refs. \cite{B,JKLW,CGC}) and dipoles
\cite{AM94,K,IM031}. Furthermore, the high--energy QCD evolution in
the leading logarithmic approximation consists indeed in the
emission of a single $s$--channel gluon at each step of the
evolution. Note that, unlike other models in the literature
\cite{Stasto05,SX05,KozLev06,BMMSX06}, our model does not include a
mechanism for particle recombination, in agreement with the fact
that there is no ($s$--channel) gluon, or dipole, recombination in
QCD in the leading--logarithmic approximation. Rather, as we shall
see, particle saturation occurs via high--density effects in the
emission rate, similar to the gluon saturation in the framework of
the JIMWLK evolution \cite{SAT}.

In zero (transverse) dimensions, that is, for a model living at a
single lattice site, the assumptions \texttt{(i-ii)} above are
sufficient to completely fix --- via the condition (\ref{eq-LorInv})
for boost invariance --- the structure of the evolution equation for
the probabilities $P$ \cite{AMSalam95,BIT06}. But with one spatial
dimension at our hand, there is still much freedom left, because one
can make different choices for the non--locality of the splitting
process. As we shall see in what follows, a relatively simple model
which is conceptually interesting (in the sense of bearing many
similarities to QCD) can be obtained after also imposing the
following, third, constraint:

\vspace*{.2cm}
\begin{enumerate}
\item[\tt(iii)]{\em After one step in the evolution, the configuration
of the system changes only by the addition of one new particle at
some arbitrary site.} That is, the configuration remains the same as
prior to the evolution except for the presence of the additional
particle. This is clearly a stronger constraint than the above
assumption \texttt{(ii)} which would allow for any final state with
one additional particle as compared to the original state,
irrespective of the spatial distribution of the particles in the
final state.

\end{enumerate}
\vspace*{.2cm}

\noindent Under this additional assumption, the evolution equation
for $P(\coll{n},Y)$ (the `master equation') takes the following
generic form:
  \beq\label{eq-dpdy}\hspace*{-.3cm}
   \frac{\del P(\coll{n},Y)}{\del Y}
   = \sum_{i}\big[
   f_i(\dots,n_i-1,\dots)\, P(\dots,n_i-1,\dots,Y)
   -f_i(\coll{n})\, P(\coll{n},Y) \big],
  \eeq
where the quantity $f_i({\coll{n}})$ has the meaning of a `deposit'
rate: this is the probability per unit rapidity to find an extra
particle at site $i$ after one step of the evolution, starting from
an original configuration $\coll{n}$. The positive (gain) term in
the r.h.s. Eq.~(\ref{eq-dpdy}) shows that in order to arrive at a
final configuration $\coll{n}$ after one step in this evolution, one
needs to start in a configuration containing one particle less on
the site $i$, where $i$ is arbitrary. The negative (loss) term is
necessary for the conservation of the probability.

The evolution law in Eq.~(\ref{eq-dpdy}) deserves some physical
discussion. But let us first complete the construction of the model,
by deducing an explicit form for the deposit rate $f_i(\coll{n})$.
As anticipated, this is obtained by exploiting the condition of
boost--invariance, that is, by inserting the generic master equation
(\ref{eq-dpdy}) into Eq.~(\ref{eq-LorInv}), to deduce the following
constraint :
  \beq\label{eq-fconst}
   \sum_i\left[
   f_i(\coll{n})\, t_i(\coll{m})-
   f_i(\coll{m})\, t_i(\coll{n})
   \right] =0,
  \eeq
where we have defined
  \beq\label{eq-ti}
   t_i(\coll{n}) = 1 - \prod_j \sigma_{ij}^{n_j}.
  \eeq
The above notation is not accidental, since $t_i(\coll{n})$ is
formally equal to the amplitude for the scattering of a projectile
particle at site $i$ off a target with a given configuration
$\coll{n}$. The most general solution to Eq.~(\ref{eq-fconst}) that
we have found reads
  \beq\label{eq-figen}
    f_i(\coll{n}) = \sum_j c_{ij} \,t_j(\coll{n})
    \qquad \text{with} \qquad c_{ij}=c_{ji},
  \eeq
and where $c_{ij}$ is independent of the particle occupation
numbers. It is convenient to choose $c_{ij} \propto \delta_{ij}$,
since this leads to a relatively simple and intuitive model which
still has a rather rich structure, as we shall later see. Namely, we
shall fix our model by choosing
  \beq\label{eq-cij}
   c_{ij} = \frac{\Delta}{\tau}\, \delta_{ij}\,,
  \eeq
where $\tau \equiv \tau_{ii}$ and $\Delta$ stands for the ``lattice
spacing'' along the discretized spatial axis. The fact that $c_{ij}$
is proportional to $\Delta$ is natural, since the deposit rate
density $f_i/\Delta$ will be well--defined in the continuum limit
$\Delta \to 0$. The fact that we choose $c_{ij}$ proportional to
$1/\tau$ is again natural, since in the low density limit $f_i$
should be of order $\order{1}$ when measured in units of $\tau$.
Indeed, when the $n_j$'s are relatively small, such that
$n_j\tau_{ij}\ll 1$, the scattering amplitude $t_i$ in
Eq.~(\ref{eq-ti}) reduces to $t_i\approx\sum_{j}\tau_{ij} n_j$ and
therefore the particular parametric dependence of $c_{ij}$ on $\tau$
exhibited above is clearly necessary for the condition to be
fulfilled.

To conclude, as a result of the various assumptions presented above,
we finally arrive at the model defined by the master equation
(\ref{eq-dpdy}) with the following expression for the deposit rate :
  \beq\label{eq-fi}
   \frac{f_i(\coll{n})}{\Delta} =
   \frac{1 - \prod_j \sigma_{ij}^{n_j}}{\tau} =
   \frac{t_i(\coll{n})}{\tau}\,.
  \eeq
Up to an overall normalization, this is simply the scattering
amplitude for the scattering between the particle newly produced at
site $i$ and the preexisting particles in the system, from which the
new particle has been emitted.

To better appreciate the physical content of the above formula, we
need to know a little bit more about the elementary scattering
amplitude $\tau_{ij}$. Later on, we shall present an explicit model
for this quantity, dictated by the analogy with QCD (see Sect.
\ref{sect-analytics} for details). Here, it suffices to mention that
$\tau_{ij}$ depends only upon the separation $|i-j|$ and is rapidly
decreasing when increasing this separation. In particular, the
quantity $\tau \equiv \tau_{ii}$ is independent of $i$ (a property
that has been already used in writing Eq.~(\ref{eq-cij})) and will
be taken to be small: $\tau \ll 1$ (the analogous quantity in QCD is
of $\order{\alpha_s^2}$). One therefore has $\tau_{ij}\ll 1$ for any
pair $(ij)$.

Armed with this knowledge, we now return to Eq.~(\ref{eq-fi}) and
discuss some limiting cases:

\texttt{(i)} When the occupation numbers are relatively small, such
that $n_j \ll 1/\tau_{ij}$, one can expand $\sigma_{ij}^{n_j}
\approx 1 - {n_j}\tau_{ij}$, and then the deposit rate density for
the given site $i$ becomes simply proportional to the particle
occupation numbers at all sites:
  \beq\label{eq-filow}
   \frac{f_i(\coll{n})}{\Delta} \,\approx\,
   \sum_j \frac{\tau_{ij}}{\tau}\,n_j\,\,
   \quad\mbox{when\, $n_j \ll 1/\tau_{ij}$ \, for\,
   any \,$j$}.
  \eeq
This describes a situation where the extra particle at $i$ is {\em
incoherently} emitted by any of the preexisting particles in the
system. When the approximation (\ref{eq-filow}) holds for {\em any}
site $i$, we shall speak about the {\em dilute}, or {\em linear},
regime. Since, as mentioned before, $\tau_{ij}$ is a decreasing
function of the separation $|i-j|$, the dilute regime is realized
when {\em all} the occupation numbers obey the condition $n_j \ll
1/\tau$.  The dipole picture in QCD \cite{AM94} is the right term of
comparison for our toy model in this dilute regime; and, indeed, if
one computes the analogous deposit rate in the dipole picture, one
finds that this is linear in the dipole density \cite{IM031}.

\texttt{(ii)} When at least one of the occupation numbers $n_j$
becomes so large that $\sigma_{ij}^{n_j}\ll 1$ (this requires $n_j
\simge 1/\tau_{ij}$), the deposit rate at $i$ saturates at its
maximal value of $\order{1/\tau}$ :
 \beq\label{eq-fihigh}
   \frac{f_i(\coll{n})}{\Delta} \,\approx\,
   \frac{1}{\tau}\,\quad\mbox{when\, $n_j \simge 1/\tau_{ij}$ \, for\,
   some \,$j$}.
  \eeq
In particular, a sufficient condition to have saturation at a given
site $i$ is that the respective occupation number $n_i$ obeys $n_i
\simge 1/\tau$. This is again similar to the situation in QCD, where
the gluon emission rate in the JIMWLK equation (the analog of the
deposit rate in the toy model) saturates when the gluon occupation
numbers are sufficiently
high. 

One may also notice that in the limit where the scattering is truly
local, $\tau_{ij} = \tau \delta_{ij}$, Eq.~(\ref{eq-fi}) yields
$f_i/\Delta=(1-\sigma^{n_i})/\tau$, with $\sigma\equiv 1-\tau$. In
this limit, the different lattice sites decouple from each other,
and each of them evolves according to the zero--dimensional model of
Refs. \cite{AMSalam95,BIT06}. This limit is uninteresting for the
present purposes, and shall not be considered anymore.

Let us now return, as promised, to a physical discussion of the
evolution described by Eq.~(\ref{eq-dpdy}). To that aim, it is
useful to imagine a specific microscopic mechanism leading to this
evolution. It is natural to assume that the process leading to the
production of the extra particle in one evolution step is the
splitting of one particle into two new ones. For this mechanism to
be consistent with Eq.~(\ref{eq-dpdy}), one additional assumption is
however needed: at least one of the two daughter particles produced
after such a splitting must remain on the same lattice site as its
parent particle. Introducing the emission rate $g_{j \to
ji}(\coll{n})$ for a particle at $j$ to split into two particles at
$j$ and $i$, respectively, with arbitrary $i$, we are lead to the
following master equation
 \beq\label{eq-dpdytemp}
  \hspace{-0.35cm}
   \frac{\del P(\coll{n},Y)}{\del Y}
   = \sum_{ij}\big[
   g_{j \to ji}(\dots,n_i-1,\dots)\, P(\dots,n_i-1,\dots,Y)
   -g_{j \to ji}(\coll{n})\, P(\coll{n},Y) \big],\nn
  \eeq
which leads to Eq.~(\ref{eq-dpdy}) after identifying  $f_i(\coll{n})
= \sum_{j} g_{j \to ji}(\coll{n})$. The process in
Eq.~(\ref{eq-dpdytemp}) is quite peculiar in the sense that the
particle which splits does not disappear in the final state, rather
it gets replaced by a new particle at the same
site\footnote{Incidentally, this discussion also shows that a
natural generalization of the present model would be to allow for
the more general splitting $i \to jk$ with generic $i$, $j$, and
$k$, and with a rate $g_{i \to jk}(\coll{n})$.}. In the analogy with
QCD --- in which, we recall, particle positions correspond to dipole
sizes
---, this would correspond to a process in which a dipole of size
$j$ splits into two dipoles, one with the same size $j$ and the
other one with some arbitrary size $i$. This is quite different from
the (BFKL) picture of dipole splitting in QCD at large $N_c$
\cite{AM94}, where the daughter dipoles have generic sizes, which
typically are comparable to the size of the parent dipole. But in
spite of this explicit dissimilarity at the level of the physical
interpretation, the toy model evolution generated by
Eqs.~(\ref{eq-dpdy}) and (\ref{eq-fi}) turns out to have many common
features with the high--energy evolution in QCD, as it should become
clear from the analysis in the subsequent sections.

Last, but not least, we notice that one can put the model in the
Hamiltonian form
  \beq\label{eq-ham}
   \frac{\del P(\coll{n},Y)}{\del Y}
   &&= - \frac{\Delta}{\tau} \,\sum_i
   \left[ 1 - \exp \left(-\frac{\del}{\del n_i}\right) \right]
   \bigg[ 1- \exp\bigg(\sum_j n_j \ln \sigma_{ij}  \bigg) \bigg]
   P(\coll{n},Y)
   \nn
   &&\equiv H\, P(\coll{n},Y),
  \eeq
where the differential operator $\exp \left(-{\del}/{\del
n_i}\right)$ is the `translation' operator which reduces the
occupation number at site $i$ by one: $\exp \left(-{\del}/{\del
n_i}\right) F(\dots,n_i,\dots)= F(\dots,n_i-1,\dots)$ for a generic
function $F(\coll{n})$. This rewriting makes it explicit that the
Hamiltonian underlying the evolution of the model is `self--dual',
that is, it is invariant under the self--duality transformation
\cite{KL3} which in the present context consists in exchanging
 \beq\label{eq-selfdual}
 \frac {\del }{\del n_i}\ \longleftrightarrow \ -\sum_j n_j \ln \sigma_{ij}\,,
 \eeq
and then reversing the order of the operators. This self--duality is
the expression of the constraint of boost--invariance on the
structure of the evolution Hamiltonian \cite{KL3,BIIT05}. Thus, this
condition is automatically fulfilled by our model, where
boost--invariance is built-in. Also, the presence of two types of
exponentials
--- one involving the particle occupation number and the other one,
the derivative with respect to it --- in Eq.~(\ref{eq-ham}) is
reminiscent of the two types of Wilson lines which appear in the QCD
Hamiltonian proposed in Refs. \cite{BREM,Balit05}. In that case,
$\sum_j n_j |\ln \sigma_{ij}|$ is replaced by the color field
produced by the $s$--channel gluons, and ${\del }/{\del n_i}$ by the
(functional) derivative with respect to the color charge density of
these gluons.

It is furthermore interesting to consider the approximate version of
the master equation which is formally obtained by expanding the
operator $\exp \left(-{\del}/{\del n_i}\right)$ to linear order in
the derivative:
  \beq\label{eq-JIMWLK}
   \frac{\del P(\coll{n},Y)}{\del Y}
   \approx - \,\sum_i
   \frac{\del}{\del n_i}\,
   \Big[f_i(\coll{n})\,P(\coll{n},Y)\Big].
   \eeq
Less formally, this is simply the limit in which the difference
between the two terms in the r.h.s. of Eq.~(\ref{eq-dpdy}) is
assimilated to a derivative with respect to $n_i$, which is
reasonable when $n_i\gg 1$, for any $i$. This is the limit of {\em
large occupation numbers}, in which the fluctuations associated with
the discreteness of the particle number become negligible.

In this limit, the self--duality of the evolution is of course lost,
since the system property of being dense depends upon the frame.
Moreover, it should be intuitively clear that, whatever frame we
choose, the system cannot be dense at {\em all} the sites. Indeed,
the spatial axis being infinite, there will always be an infinite
number of sites where the occupation numbers are either zero, or
small, of $\order{1}$, and this for any $Y$. Hence,
Eq.~(\ref{eq-JIMWLK}) is merely a formal approximation, which for a
given $Y$ has, at most, a limited range of validity in $x$ --- it
refers to the evolution of the {\em bulk} of the particles in a
region with high occupancy.

This discussion is reminiscent of that of the JIMWLK equation in
QCD, which has been established \cite{JKLW,CGC} for systems with a
relatively high gluon density and which ignores the fluctuations in
the gluon number \cite{IT04}. And, indeed, one can recognize the
above equation (\ref{eq-JIMWLK}) as the toy--model version of the
JIMWLK equation. This analogy has been already discussed in the
context of the zero--dimensional model in Ref. \cite{BIT06}, to
which we refer for more details. To summarize this section, the
toy--model that we have here introduced reduces to (one--dimensional
versions of the) dipole picture in the dilute regime and,
respectively, to the JIMWLK picture in the regime of high occupancy,
and in general it provides a self--dual `interpolation' between
these two pictures, as necessary for boost--invariance.

\section{Evolution equations for the observables}\label{sect-observ}
\setcounter{equation}{0}

Given the master equation (\ref{eq-dpdy}) along with the deposit
rate, Eq.~(\ref{eq-fi}), one can obtain the evolution equation for
any `observable' $\mcal{O}$, by which we shall mean a quantity whose
event--by--event value $\mcal{O}(\coll{n})$ depends upon the
configuration $\coll{n}$ of the particles in the system. Its average
value at rapidity $Y$, which is a measurable quantity, is given by
  \beq\label{eq-Oave}
   \avg{\mcal{O}}_Y = \sum_{\coll{n}}
   P(\coll{n},Y)\, \mcal{O}(\coll{n}).
  \eeq
Differentiating with respect to $Y$ and performing a shift $n_i-1
\to n_i $ in the contributions arising from the gain (plus sign)
terms in Eq.~(\ref{eq-dpdy}) one easily finds
  \beq\label{eq-Oevol}
   \frac{\del \avg{\mcal{O}}_Y}{\del Y} =
   \sum_i \big\langle f_i(\coll{n})\,
   \big[\mcal{O}(\dots,n_i +1,\dots) - \mcal{O}(\coll{n})\big
   ]\big\rangle_Y.
  \eeq
In what follows, we shall specialize the generic evolution equation
(\ref{eq-Oevol}) to two interesting types of observables, namely,
the $k$--body particle occupation numbers and the scattering
amplitudes for projectiles made with a given number of particles. In
the process, we shall also demonstrate that the continuum limit
$\Delta\to 0$ of the ensuing equations is well defined, and we shall
furthermore compare these equations to the corresponding ones in
QCD.

\subsection{Particle number densities}

The simplest observable that one can think of is $\avg{n_i}_Y$ ---
the average occupation number at the site $i$. From
Eq.~(\ref{eq-Oevol}), it is straightforward to obtain
  \beq\label{eq-dndy}
   \frac{\del \avg{n_i}_Y}{\del Y} =
   \avg{\frac{\Delta (1- \prod_j \sigma_{ij}^{n_j})}{\tau}}_Y =
   \avg{f_i(\coll{n})}_Y,
  \eeq
with an obvious physical interpretation: the rate for the change in
the average particle number at a given site is equal to the average
value of the deposit rate at that particular site. Two limiting
cases of Eq.~(\ref{eq-dndy}) are particularly interesting,
corresponding to the two limits of the deposit rate $f_i$ discussed
in relation with Eq.~(\ref{eq-filow}) and, respectively,
Eq.~(\ref{eq-fihigh}) :

\texttt{(i)} If the system is relatively dilute `around the site
$i$' --- meaning that, for the typical configurations, one has $n_i
\ll 1/\tau$ and also $n_j \ll 1/\tau$ for all the sites $j$ which
are not too far away from $i$ ---, then we can linearize
$\avg{f_i}_Y$ with respect to the (average) occupation numbers and
thus obtain:
 \beq\label{eq-dndylow}
   \frac{\del \avg{n_i}_Y}{\del Y} \,\approx\,
   \frac{\Delta}{\tau}\, \sum_j \tau_{ij}\avg{n_j}_Y\,.
  \eeq
In this regime, the particle occupation at site $i$ and at the
neighboring sites undergoes a {\em linear} evolution, leading to a
rapid (exponential in $Y$) rise in the respective average occupation
numbers. This rapid rise, which will be explicitly demonstrated in
Sect. \ref{sect-analytics}, is the toy--model analog of the BFKL
growth in the gluon (or dipole) distribution in QCD in the dilute
regime \cite{BFKL}.

\texttt{(ii)}  If, on the other hand, the system is `dense at $i$',
meaning that the average deposit rate $\avg{f_i}_Y$ saturates at its
maximum value, cf. Eq.~(\ref{eq-fihigh}), then the growth of the
(average) occupation number at $i$ is considerably slowed down:
  \beq\label{eq-dndyhigh}
   \frac{\del \avg{n_i}_Y}{\del Y} \,\approx\,
   \frac{\Delta}{\tau}\, .
  \eeq
This equation shows that, once the particle occupation number at $i$
becomes of $\order{1/\tau}$, its subsequent growth with $Y$ is only
very slow, {\em linear} in $Y$. This is the mechanism of {\em
saturation} within the toy model, and is similar to gluon saturation
in QCD, where the occupation numbers for the saturated modes grow
also linearly with $Y$ \cite{AM99,SAT,GAUSS}.

Returning to Eq.~(\ref{eq-dndy}), one should also notice that this
is not a closed equation for $\avg{n_i}_Y$, but rather the beginning
of a {\em hierarchy} : if one expands $\sigma_{ij}^{n_j} =
\exp\{n_j\ln\sigma_{ij}\}$ in powers of $n_j$ inside the brackets in
the r.h.s., one generates $k$--body correlations of the occupation
numbers with any $k\ge 1$. This invites us to consider the evolution
equations obeyed by the general, `normal--ordered', $k$--body
occupation numbers, defined as
  \beq\label{eq-nk}
   n^{(k)}_{i_1 \dots i_k} \equiv
   n_{i_1} (n_{i_2} - \delta_{i_1 i_2}) \dots
   (n_{i_k} - \delta_{i_1 i_k} - \dots -\delta_{i_{k-1} i_k}).
  \eeq
Here, the `normal ordering' refers to the subtraction of
$\delta$--functions, as explicit in the equation above, which are
needed to guarantee that, in constructing $n^{(k)}$, one counts only
sets of $k$ particles which are all different from each other.  The
corresponding evolution equation can be readily obtained as
  \beq\label{eq-dnkdy}
   \frac{\del \avg{n^{(k)}_{i_1 i_2 \dots i_k}}_Y}{\del Y} =
   \sum_{j = 1}^k \avg{f_{i_j}(\coll{n})\, n^{(k-1)}_{i_1 i_2 \slashed{i_j}
   \dots
   i_k}}_Y,
  \eeq
where the notation $\slashed{i_j}$ means that the $i_j$ variable is
to be omitted.

Before proceeding with the hierarchy for the scattering amplitudes,
let us first briefly discuss the {\em continuum limit} of our model,
which is straightforward. Clearly, this limit amounts to replacing
$i \to x_i= i \Delta$ and then letting $\Delta \to 0$, so that
$x_i\to x$, with $x$ our continuous spatial variable. Then, a sum
over $i$ gets converted into an integration over $x$:
  \beq\label{sumtoint}
   \Delta \sum_i F_i \to \int \dif x\, F(x),
  \eeq
while a sum over all the possible configurations of the onium
wavefunction becomes a path integral, that is
  \beq\label{sumtopath}
   \sum_{\coll{n}} \to \int [D n(x)].
  \eeq
The $k$--body (normal--ordered) particle {\em densities} are
obtained from the corresponding occupation numbers after dividing by
$\Delta^k$ and then letting $\Delta \to 0$. For example, the
2--body, or pair, density is given by
  \beq\label{eq-n2}
   n^{(2)}(x,y) = \lim_{\Delta \to 0}
   \left(
   \frac{n_i n_j}{\Delta^2}
   - \frac{\delta_{ij}}{\Delta}\,\frac{n_i}{\Delta}
   \right) =
   n(x)\,n(y) -\delta(x-y)\, n(x),
  \eeq
while the deposit rate density can be immediately obtained from
Eq.~(\ref{eq-fi}), and reads
  \beq\label{eq-fx}
   f(x) = \lim_{\Delta \to 0}\frac{f_i(\coll{n})}{\Delta} \, = \,
   \frac{1- \exp \left[
   \int \dif z\, n(z) \ln \sigma(x|z) \right]}{\tau} \,\equiv\,
   \frac{t(x)}{\tau}.
  \eeq
It is now straightforward to reexpress all the equations presented
earlier in continuum notations. For instance, the 1--body and
2--body particle densities obey the following equations, as obtained
by taking the continuum limit in Eqs.~(\ref{eq-dndy}) and
(\ref{eq-dnkdy}) (with $k=2$), respectively:
 \beq\label{eq-dndycont}
   \frac{\del \avg{n(x)}_Y}{\del Y} &=&
   \avg{f(x)}_Y,\\
   \label{eq-dndy2cont}
\frac{\del \avg{n^{(2)}(x,y)}_Y}{\del Y} &=&
   \avg{f(x)n(y)+f(y)n(x)}_Y.
  \eeq
It is important to notice that Eq.~(\ref{eq-dndy2cont}) has no
singularities in the equal--point limit $y=x$ thanks to the
subtraction of the $\delta$--function in the definition
(\ref{eq-n2}) of the normal--ordered 2--body density. A similar
property holds for the $k$--body densities with $k>2$.


\subsection{Scattering amplitudes: Beyond the Balitsky--JIMWLK equations}
\label{sect-scatt}

We now move to the scattering problem and start by assuming that the
projectile consists in a single particle located at $i$. In QCD,
this would correspond to a projectile made with a single dipole of
size $r_i\propto\exp(-x_i/2)$. This implicitly means that we work in
a frame in which almost all of the rapidity $Y$ is carried by the
target, whose wavefunction evolves according to Eqs.~(\ref{eq-dpdy})
and (\ref{eq-fi}), whereas the rapidity of the projectile is so low
that its evolution can be neglected. The $S$--matrix describing this
scattering is obtained by replacing $m_j \to \delta_{jk}$ in
Eq.~(\ref{eq-sconf}), which then yields $s_k(\coll{n}) = \prod_l
\sigma_{lk}^{n_l}$. By making use of Eq.~(\ref{eq-Oevol}), we deduce
the following evolution equation
  \beq\label{eq-dsdy}
   \frac{\del \avg{s_i}_Y}{\del Y} =
   \sum_j \frac{\Delta (1-\sigma_{ij})}{\tau}
   \,\avg{s_i s_j -s_i}_Y\,,
  \eeq
which is not a closed equation --- the average $S$--matrix element
for the one--particle projectile being related to that for a
projectile made with two particles ---, but only the first equation
in an infinite hierarchy. The general equation in this hierarchy can
be obtained by studying the scattering of a projectile made with $m$
particles, at given positions\footnote{Of course, some of these $m$
particles can have identical positions.} $i_1,i_2,...,i_m$. For a
given target configuration $\coll{n}$, the corresponding $S$--matrix
reads
  \beq\label{eq-sm}
   s_{i_1}\dots s_{i_m} =
   \prod_{j_1} \sigma_{i_1 j_1}^{n_{j_1}}\dots
   \prod_{j_m} \sigma_{i_m j_m}^{n_{j_m}} =
   \prod_j \left( \sigma_{i_1 j} \dots \sigma_{i_m j} \right)^{n_j}
  \eeq
and then Eq.~(\ref{eq-Oevol}) implies
  \beq\label{eq-dsmdy}
   \frac{\del \avg{s_{i_1}\dots s_{i_m}}_Y}{\del Y} =
   \sum_j \frac{\Delta (1- \sigma_{i_1 j}\dots \sigma_{i_m j})}{\tau}
   \, \avg{s_{i_1}\dots s_{i_m} s_j - s_{i_1}\dots s_{i_m}}_Y.
  \eeq
Although obtained here by following the evolution of the target,
these equations can be easily reinterpreted as describing evolution
in the projectile. To that aim, it is important to notice that the
kernel in the above equation, i.e. the quantity in front of the
expectation value in the r.h.s., is precisely $f_j(\coll{m})$, that
is, the deposit rate at site $j$ evaluated for the given
configuration $\coll{m}$ of the {\em projectile}. Thus,
Eq.~(\ref{eq-dsmdy}) can be interpreted as follows: a splitting
takes place in the projectile, leading to a system with $m+1$
particles (with $j$ being the position of the extra particle).
Subsequently this new system scatters off the target giving rise to
the first term in the above equation, while the second (negative
sign) term there corresponds to the possibility that no extra
particle was created.

The physical content of this hierarchy will be further discussed on
its continuum version, which reads (cf. Eq.~(\ref{eq-fi}))
  \beq\label{eq-dsmxdy}
   \frac{\del \avg{s_{x_1}\dots s_{x_m}}}{\del Y} =
   \int\limits_z f_z(\coll{m})
   \, \avg{s_{x_1}\dots s_{x_m} s_z - s_{x_1}\dots s_{x_m}},
  \eeq
in simplified notations where the coordinates are shown as lower
indices and the $Y$--dependence of the expectation values is not
indicated. The general structure of these equations is interesting,
as it may shed light on the corresponding structure in QCD. It is
especially instructive to compare these equations to the
Balitsky--JIMWLK equations \cite{B,K,JKLW,CGC,W} and also to the
more complete `Pomeron loop' equations recently proposed in QCD at
large $N_c$ \cite{IT04,MSW05,IT05}.

Recall at this point that the Balitsky hierarchy in QCD has been
obtained by performing {\em different} evolutions in the projectile
and the target: the projectile has been assumed to be dilute and
evolve according to the dipole picture\footnote{We here restrict
ourselves to the large--$N_c$ version of the Balitsky hierarchy, as
appropriate for comparison with the toy model.}, whereas the target
was taken to be dense and evolve according to JIMWLK equation. Under
these assumptions, the Balitsky equations are {\em formally} boost
invariant, but the whole scheme is clearly incomplete and
unsatisfactory, since an evolving projectile eventually becomes
dense and, vice versa, even a target which looks dense on some
resolution scale (at relatively low transverse momenta) has
necessarily a dilute tail at high transverse momenta, and this tail
is the driving force for its evolution \cite{IMM04}.

This problem is overcome, by construction, within the toy model,
where the target and the projectile are symmetrically treated, and
it is interesting to see how this is reflected in the structure of
the evolution equations. The first equation in the hierarchy
(\ref{eq-dsmxdy}), that is,
  \beq\label{eq-dsxdy}
   \frac{\del \avg{s_x}}{\del Y} =
   \int\limits_z \frac{\tau_{xz}}{\tau}
   \avg{s_x s_z - s_x},
  \eeq
is formally similar to the corresponding Balitsky equation, from
which it differs only by the replacement of the `dipole kernel' (the
rate for dipole splitting in QCD) by the reduced scattering
amplitude ${\tau_{xz}}/{\tau}$, which plays the role of the
elementary splitting rate (corresponding to a parent particle in
isolation) within the toy model. Note that the splitting is more
constrained in the toy model than in QCD, for the reasons explained
in Sect. \ref{sect-const}: one of daughter `dipoles' $x$ and $z$
which appear in the r.h.s. of Eq.~(\ref{eq-dsxdy}) is bound to have
the same `size' $x$ as its parent `dipole' in the l.h.s.

It seems therefore natural to identify the elementary splitting
$x\to xz$ with rate ${\tau_{xz}}/{\tau}$ in the toy model to the
dipole splitting in QCD. With this identification, the differences
between the toy--model hierarchy (\ref{eq-dsmxdy}) and the Balitsky
equations become visible in the higher equations in the hierarchy,
starting with the second one:
  \beq\label{eq-ds2dy}
   \frac{\del \avg{s_x s_y}}{\del Y} =
   \int\limits_z \frac{\tau_{xz} + \tau_{zy} - \tau_{xz}\tau_{zy}}
   {\tau} \avg{s_x s_y s_z - s_x s_y}.
  \eeq
The respective Balitsky equation would involve only the two positive
terms, $\tau_{xz}$ and $\tau_{zy}$,  in the numerator in the kernel,
which describe the independent splittings of the particle at $x$ and
at $y$, respectively. The additional, negative, term
$-\tau_{xz}\tau_{zy}$ corresponds to `saturation effects' in the
evolution of the projectile, namely to the fact that the two
splittings are not truly independent. This term is formally
suppressed by a power of $\tau$ with respect to the previous ones,
but as we shall shortly argue it cannot be neglected, since it plays
an essential role in the evolution.

This discussion can be easily generalized to the higher equations in
the hierarchy: the toy--model analog of the Balitsky equations can
be obtained from the general equations (\ref{eq-dsmxdy}) by
replacing the kernel $f_z(\coll{m})$ there by its linearized version
(cf. Eqs.~(\ref{eq-filow}) and (\ref{eq-fx}))
 \beq
 f_z(\coll{m})\,\longrightarrow\,\sum_{i=1}^m
  \frac{\tau(x_i|z)}{\tau}\,,
  \eeq
which describes the independent splitting $x_i\to x_iz$ of any of
the $m$ particles in the projectile. Clearly, this is similar to
QCD, since it is tantamount to saying that the projectile evolves
according to the dilute approximation (\ref{eq-filow}) to the master
equation (\ref{eq-dpdy}). Also like in QCD, it can be easily checked
that the same set of equations would be obtained by evolving the
target according to the toy--model version of the `JIMWLK' equation,
Eq.~(\ref{eq-JIMWLK}).

Thus, clearly, the toy--model hierarchy (\ref{eq-dsmxdy}) goes
beyond the Balitsky--JIMWLK equations by including saturation
effects in the evolution of the projectile, or, equivalently,
particle--number fluctuations in that of the target. It is therefore
interesting to better understand the physical role of the additional
term, so like the negative term in the kernel in
Eq.~(\ref{eq-ds2dy}). To that aim, and also to facilitate the
comparison with the `Pomeron loop' equations of QCD, it is
convenient to rewrite the above equations in terms of the
$T$--matrix elements (or `scattering amplitudes') $t_x\equiv 1-s_x$.
We display here the first two equations in the ensuing hierarchy
(the third such equation will be shown in the Appendix):
 \beq\label{eq-dtxdy}
   \frac{\del \avg{t_x}}{\del Y} =
   \int\limits_z \frac{\tau_{xz}}{\tau}
   \avg{t_z - t_x t_z},
  \eeq
and, respectively,
  \beq\label{eq-dt2dy}
   \hspace{-0.55cm}
   \frac{\del \avg{t_x t_y}}{\del Y} =
   \int\limits_z \left[
   \frac{\tau_{xz}}{\tau} \avg{(t_z - t_x t_z) t_y}+
   \frac{\tau_{yz}}{\tau} \avg{(t_z - t_y t_z) t_x}+
   \frac{\tau_{xz} \tau_{yz}}{\tau} \avg{t_z (1-t_x)(1-t_y)}
   \right]\!.
  \eeq
Notice that $t_x t_y=(1-s_x)(1-s_y)$ is the amplitude for the {\em
simultaneous} scattering of two particles in the projectile. At this
level, it becomes straightforward to recognize the toy--model
analogs of the BFKL and, respectively, BK equation: the former is
obtained by neglecting the non--linear term $\avg{t_x t_z}$ in the
r.h.s. of Eq.~(\ref{eq-dtxdy}), while the second corresponds to
treating this term in a mean--field approximation (MFA) which
assumes factorization: $\avg{t_x t_z} \approx\avg{t_x} \avg{t_z}$.
However, this factorization is inconsistent with the higher
equations in the hierarchy, starting with Eq.~(\ref{eq-dt2dy}). This
breakdown of the MFA is associated with the presence of `fluctuation
terms', like the last term $\propto \tau_{xz}\tau_{zy}$ in the
r.h.s. of Eq.~(\ref{eq-dt2dy}), which are precisely the additional
terms with respect to the (toy--model) Balitsky equations. Such
terms are important, since they generate the correlations in the
dilute regime, as we explain now:

Consider indeed the regime in which the target is so dilute that the
average scattering amplitude is {\em very} weak\footnote{The
relevance of this regime for the high--energy evolution will be
explained in Sects. \ref{sect-analytics} and \ref{sect-numerics}.}:
$\avg{t} \lesssim \tau$. This corresponds to the dilute tail of the
target distribution at large values of $x$, where the (average)
occupation number is $\avg{n} \lesssim 1$. (Indeed, in the dilute
regime, one can write $\avg{t_x}\approx \int_z \tau_{xz} \avg{n_z}\,
$; see, e.g., Eq.~(\ref{eq-ti}).) In this regime, the last term in
the r.h.s. of Eq.~(\ref{eq-dt2dy}) can be approximated as
$({\tau_{xz} \tau_{yz}}/{\tau}) \avg{t_z}$, which for $\avg{t}
\lesssim \tau$ is comparable to, or even larger than, the BFKL--like
terms in that equation, so like $({\tau_{xz}}/{\tau}) \avg{t_z
t_y}$. This shows that, in this dilute regime, the two--particle
amplitude $\avg{t t}$ is still very small but it gets built from the
one--particle amplitude $\avg{t}$, via the last term in
Eq.~(\ref{eq-dt2dy}). To better appreciate the physical
interpretation of this term, notice that in the dilute regime one
can successively write
 \beq\label{eq-fluct}
  \frac{\del \avg{t_x t_y}}{\del Y}\,
    \bigg|_{\rm fluct} \,\simeq\,
   \int\limits_z
   \frac{\tau_{xz} \tau_{yz}}{\tau} \avg{t_z}\,\simeq\,
 \int\limits_z
   \frac{\tau_{xz} \tau_{yz}}{\tau} \int\limits_w \tau_{zw} \avg{n_w}
  \,\simeq\,
 \int\limits_z \tau_{xz} \tau_{yz}\,
   \frac{\del \avg{n_z}}{\del Y},
  \eeq
where we have also used the appropriate limit of
Eq.~(\ref{eq-dndycont}). According to this equation, the change in
the two--particle amplitude $\avg{t_x t_y}$ in one step of the
evolution can be interpreted as the following evolution in the
target: first, a new particle is created at $z$, via the splitting
process $w\to wz$, with any $w$; then, the new target particle at
$z$  scatters simultaneously with the two projectile particles at
$x$ and $y$.

We see that the two--particle correlation $\avg{t_x
t_y}-\avg{t_x}\avg{t_y}$ gets built from `target fluctuations'
(i.e., splitting processes leading to a change in the particle
number in the target distribution), via {\em multiple scattering}.
This mechanism is quite similar to that identified in the context of
QCD in Ref. \cite{IT04}, and which has inspired an extension of the
Balitsky--JIMWLK hierarchy known as the `Pomeron loop' hierarchy
\cite{IT04,MSW05,IT05}. The general structure of this `Pomeron loop'
hierarchy is indeed similar to that for the scattering amplitudes in
the toy model (cf. Eqs.~(\ref{eq-dtxdy})--(\ref{eq-dt2dy})), in the
sense of including BFKL terms, non--linear terms responsible for
unitarization, and fluctuation terms which generate correlations in
the dilute regime. However, some subtle differences persist, which
can be partly attributed to real structural differences between the
dynamics in QCD and that in the toy model, and partly to the
different ways how perturbation theory is organized in the two
cases. It turns out that it is instructive to understand these
differences in more detail, as they shed new light on the QCD
equations themselves. This will be discussed at length in the
Appendix, where we shall see the two sets of equations are
essentially equivalent at low density, but they potentially differ
from each other in their respective extensions towards the
high--density regime. In that sense, the lessons drawn from the toy
model may suggest improvements of the `Pomeron loop' equations in
QCD.

\section{Analytic results: From BFKL growth to diffusive scaling}
\label{sect-analytics} \setcounter{equation}{0}

In this section we shall present an analytic study of the evolution
described by the toy model which mimics the corresponding analysis
of the high energy evolution in QCD, as standard by now in the
literature (see especially Refs.
\cite{SCALING,MT02,MP03,IMM04,IT04,HIMST06}). We shall thus
successively address the linear approximation (which corresponds to
the BFKL equation \cite{BFKL} in QCD), the mean field approximation
(the toy--model analog of the BK equation \cite{K}), and, finally,
the effects of particle--number fluctuations, that we shall describe
by analogy\footnote{The pertinence of this analogy for the particle
model at hand will be confirmed by the numerical analysis of this
model in Sect. \ref{sect-numerics}.} with the reaction--diffusion
problem in statistical physics (so like in Ref. \cite{IMM04} for
QCD).

For the purposes of this analysis, we need to be more specific about
the form of the elementary particle--particle scattering amplitude
$\tau(x|z)$. Inspired by the analogy with QCD, in which this
quantity corresponds to the amplitude for dipole--dipole scattering,
we shall choose
  \beq\label{eq-tau}
   \tau(x|z) = \tau \exp(-|x-z|).
  \eeq
The QCD--origin of this formula can be recognized as follows:  after
translating to QCD notations, that is, $x \equiv \ln(r_0^2/r^2)$
with $r$ the current dipole size and $r_0$ an arbitrary scale of
reference, Eq.~(\ref{eq-tau}) becomes equivalent to $\tau(r_1|r_2) =
\tau\, r_<^2/r_>^2 $, where $r_< = {\rm min}(r_1,r_2)$ and $r_> =
{\rm max}(r_1,r_2)$. With the further identification $\tau =
\alpha_s^2$, the latter is a good approximation to the amplitude for
the central scattering between two elementary dipoles in QCD. By
`central' we mean, as usual, a collision at zero relative impact
parameter. It is also understood that the amplitude is averaged over
the orientations of the two--dimensional vectors $\bm{r}_1$ and
$\bm{r}_2$.

\subsection{BFKL evolution in the toy model}

We shall focus on the equations obeyed by the dipole scattering
amplitudes and start with the linearized version of
Eq.~(\ref{eq-dtxdy}), which plays the role of BFKL equation
\cite{BFKL} within the toy model. With the specific form of
$\tau_{xz}$ given above, this equation reads
  \beq\label{eq-lin}
   \frac{\del \, t_x}{\del Y} =
   \int \dif z\, \exp(-|x-z|)\, t_z
  \eeq
(averages are implicitly assumed). Defining the Mellin transform
with respect to $\exp(-x)$
  \beq\label{eq-tgamma}
   \tilde{t}(\gamma,Y) =
   \int\limits_{-\infty}^{\infty}
   \dif x\, \exp(\gamma x)\,t(x,Y),
  \eeq
one finds that $\tilde{t}(\gamma,Y) = \tilde{t}(\gamma,\!Y=\!0)
\exp[\chi(\gamma)\,Y]$ where the `characteristic function'
$\chi(\gamma)$ is the Mellin transform of the kernel in
Eq.~(\ref{eq-lin}) and reads
  \beq\label{eq-chi}
   \chi(\gamma) = \frac{1}{1-\gamma} +\frac{1}{1+\gamma}\,.
  \eeq
Then the general solution to Eq.~(\ref{eq-lin}) is obtained as
  \beq\label{eq-linsol}
   t(x,Y) =
   \int\limits_{\mcal{C}} \,\frac{\dif \gamma}{2 \pi \rmi}\,\,
   \tilde{t}(\gamma,0)\,\exp[\chi(\gamma) Y - \gamma x]\,,
  \eeq
where the integration contour $\mcal{C}$ is parallel to the
imaginary axis and with its real part being such that $|{\rm
Re}(\gamma)| <1$. From Eq.~(\ref{eq-tgamma}) it is clear that
$\tilde{t}(\gamma,0)$ is the Mellin transform of the scattering
amplitude at $Y=0$. For definiteness, we shall assume that the
target is initially composed a single dipole of size $r_0 =
\exp(-x_0/2) $, in which case we have $\tilde{t}(\gamma,0) = \tau\,
\chi(\gamma) \exp(\gamma x_0)$.

Let us now analyze some special limits of Eq.~(\ref{eq-linsol}), to
which we shall refer by using a QCD--inspired terminology:

\vspace*{.2cm}
 \noindent{\tt (i)} {\em The Pomeron intercept :}
Consider the high--energy limit $Y \to \infty$ at fixed $x$ (which
is the limit used in QCD to define the `Pomeron intercept'). In the
kinematical plane $(Y,x)$, this corresponds to the evolution along a
(nearly) vertical axis. In this limit, the integral is dominated by
the region around the point $\gamma_{\mathbb{P}}$ that satisfies
$\chi'(\gamma_{\mathbb{P}})=0$. One easily finds
$\gamma_{\mathbb{P}} = 0$ and $\chi(\gamma_{\mathbb{P}}) = 2$, and
by also performing the Gaussian integration around this saddle point
one obtains
  \beq\label{eq-pomsol}
   t(x,Y) = \frac{\tau}{\sqrt{2 \pi Y}}\,
   \exp\left[ 2 Y - \frac{(x-x_0)^2}{8 Y}\right].
  \eeq
The $Y$--dependence of the amplitude is similar to the QCD one, as
in both cases it increases exponentially in $Y$. (Recall that in QCD
the dominant $Y$--dependence in this limit is
$\sim\exp(\omega_{\mathbb{P}} Y)$, with $\omega_{\mathbb{P}} = 4 \ln
2$.) However the analogy is not complete, since the toy--model
amplitude as given above is independent\footnote{In the QCD
terminology, the high--energy solution (\ref{eq-pomsol}) exhibits an
`anomalous dimension' $1-\gamma_{\mathbb{P}} = 1$, which is the
maximal possible value; for comparison, this value is $1/2$ for the
corresponding solution to the BFKL equation.} of the projectile
dipole size $r$. By contrast, the respective amplitude in QCD is
proportional to $r$, which is however a relatively `weak' dependence
when compared to the corresponding result at fixed order
perturbation theory, namely $t(r)\sim r^2$ (`color transparency').
Note also the Gaussian dependence of Eq.~(\ref{eq-pomsol}) upon
$x-x_0$, which describes diffusion with a diffusive radius $\propto
Y^{1/2}$. This is again similar to the well--known BFKL diffusion in
the logarithmic variable $\ln(1/r^2)$.

\vspace*{.2cm} \noindent{\tt (ii)} {\em Double logarithmic
approximation (DLA) :} This is the evolution along a direction in
the plane $(Y,x)$ which is such that the difference $x-x_0$ between
the dipole sizes increases faster than the rapidity $Y$. When $x-x_0
\ggg Y$, the integral is dominated by a value of $\gamma$ which is
close to 1. The saddle point occurs at $\gamma_{\rm DLA} = 1 -
[Y/(x-x_0)]^{1/2}$ and we are lead to
  \beq\label{eq-dlasol}
   t(x,Y) = \frac{\tau}{2 \sqrt{\pi}}\,
   [(x-x_0) Y]^{-1/4}
   \exp \left[-(x-x_0) + \sqrt{4 (x-x_0) Y} \right].
  \eeq
This is exactly the same as the corresponding QCD result, which
should not come as a surprise: in the `double logarithmic limit' of
QCD, one of the daughter dipoles has a size equal to the parent one,
a feature which is a built-in in the one--dimensional model under
consideration. In particular, the above result shows `color
transparency', that is, $t(x)\propto \exp [-(x-x_0)]$, in full
analogy with the QCD behaviour $t(r)\sim r^2$ at very small values
of $r$.

\subsection{Unitarity corrections in the mean field approximation}

As explicit in the previous analysis, the high--energy evolution
described by Eq.~(\ref{eq-lin}) leads to a scattering amplitude
which rises rapidly with $Y$ (cf. Eq.~(\ref{eq-pomsol})), and thus
eventually violates the unitarity bound $t\le 1$. So, to study the
high--energy behaviour, one needs to include the non--linear terms
responsible for unitarity corrections. We shall first do so within
the {\em mean field approximation} $\avg{t_x t_y} \approx \avg{t_x}
\avg{t_y}$, in which Eq.~(\ref{eq-dtxdy}) reduces to a closed,
non--linear equation (the expectation values are again implicit)
 \beq\label{eq-linBK}
   \frac{\del \, t_x}{\del Y} =
   \int \dif z\, \exp(-|x-z|)\, (t_z - t_xt_z)
  \eeq
which is the toy model analog of the BK equation in QCD \cite{B,K}.
This analogy is not only formal, but it covers several essential
aspects: \texttt{(a)} Like the BK equation, Eq.~(\ref{eq-linBK}) is
consistent with unitarity and, moreover, its solution approaches the
unitarity bound at large $Y$ (as obvious from the fact that $t=1$ is
a fixed point). \texttt{(b)} In the weak scattering limit $t\ll 1$,
Eq.~(\ref{eq-linBK}) reduces, as we have seen, to a linear equation
which describes an exponential increase with $Y$ and diffusion in
$x$. Together, these properties imply Eq.~(\ref{eq-linBK}) is in the
universality class of the FKPP equation \cite{Saar}, so like the BK
equation itself \cite{MP03}, which in turn implies that the
evolution towards saturation is driven by the linear dynamics in the
dilute regime --- the precise form of the non--linear terms is
unimportant so long as they provide saturation. It is then easy to
adapt the analysis of the BK equation in Refs. \cite{SAT,MT02,MP03}
to Eq.~(\ref{eq-linBK}), with the following results:

For sufficiently large values of $Y$ (in order to loose memory of
the initial condition and reach a universal behaviour), the solution
$t(x,Y)$ is a {\em traveling wave}, that is, a front which
interpolates between $t=1$ at large negative values of $x$ and $t\to
0$ at large positive values and which propagates towards larger
values of $x$ when increasing $Y$. The position $x_s(Y)$ of this
front defines the {\em saturation line}, that is, the direction of
evolution in the plane $(Y,x)$ along which the scattering amplitude
is constant and of $\order{1}$. As mentioned before, the location of
this line can be inferred from the solution to the linearized
(`BFKL') equation (\ref{eq-lin}). We thus return to this solution,
Eq.~(\ref{eq-linsol}), and consider a third direction of evolution
in the $(Y,x)$--plane, that corresponding to the saturation line:

\vspace*{.2cm} \noindent{\tt (iii)} {\em The saturation line :} When
increasing the rapidity along this line, the position $x$ (i.e., the
inverse dipole size) should be correspondingly increased in order
for the amplitude to remain constant. For $Y$ large enough, one can
use the saddle point approximation in Eq.~(\ref{eq-linsol}). The
saddle point condition
 \beq\label{eq-dchisat}
   \chi'(\gamma_s)Y -x_s = 0,
   \eeq
together with the condition that the exponent vanishes (in order for
the amplitude to be roughly constant) along the saturation line
  \beq\label{eq-chisat}
   \chi(\gamma_s)Y - \gamma_s x_s = 0,
  \eeq
uniquely determine the value of the saturation saddle point
$\gamma_s$ and the line $x_s(Y)$ (the latter up to an additive
constant), which in this approximation is simply a straight line:
$x_s(Y)\approx \lambda_s Y$. Namely, one finds
  \beq\label{eq-gammas}
   \chi'(\gamma_s) = \frac{\chi(\gamma_s)}{\gamma_s}
   \Rightarrow \gamma_s = \frac{1}{\sqrt{3}}\,,\quad\mbox{and}\quad
   \lambda_s =
   \frac{\chi(\gamma_s)}{\gamma_s}   = 3 \sqrt{3}\,.
  \eeq
It is in fact possible to improve our estimate for $\lambda_s$ by
more properly taking into account the non--linear effects. (Within
the context of the linear equation, this requires introducing an
absorptive boundary parallel to the saturation line; see Ref.
\cite{MT02} for details.) One thus finds
  \beq\label{eq-lambda}
   \lambda_s \equiv \frac{\dif x_s(Y)}{\dY} \,\approx\,
   \frac{\chi(\gamma_s)}{\gamma_s} - \frac{3}{2 \gamma_s Y} =
   3 \sqrt{3} - \frac{3 \sqrt{3}}{2 Y}\,,
  \eeq
in agreement with results from the FKPP equation \cite{Saar}. To the
same accuracy, the amplitude in the vicinity of the saturation line
can be obtained via an expansion around the saturation saddle point,
which yields
   \beq\label{eq-satsol}
   t(x,Y) =
   c_1 \tau (x-x_s+c_2) \exp\left[
   -\gamma_s (x-x_s) - \frac{(x-x_s)^2}{2 \chi''(\gamma_s) Y}
   \right].
  \eeq
This is strictly valid for $1 \ll x-x_s \ll 2 \chi''(\gamma_s) Y$,
which is a parametrically large window at large $Y$. In
Eq.~(\ref{eq-satsol}), $c_1$ and $c_2$ are unknown constants of
order $\order{1}$, $\chi''(\gamma_s) = 27$, and the linear factor
$x-x_s$ arises from the absorptive boundary.

We notice that in the region $x-x_s \ll \sqrt{2 \chi''(\gamma_s)
Y}$, where the diffusion term in the exponent in
Eq.~(\ref{eq-satsol}) can be neglected, the amplitude exhibits {\em
geometric scaling}, i.e.~it depends on $x$ and $Y$ only through the
combined variable $x-x_s(Y)$. This is what one means by a `traveling
wave' \cite{MP03} : a front which when increasing $Y$ gets simply
translated towards larger values of $x$, without being distorted.

It is furthermore instructive to translate the above picture to the
average particle number in the target, $n(x,Y)$. In the tail of the
front at $x\gg x_s(Y)$, where $t\ll 1$, we have $t(x,Y)\approx \int
\dif z\, \tau(x|z) n(z,Y)$, where the $z$--integration is peaked at
$z=x$. Thus, a `small' (large--$x$) incoming `dipole' scatters off
target `dipoles' with a similar `size' ($z\sim x$), and the weakness
of the interaction corresponds to the fact that the target looks
{\em dilute} on this resolution scale: $n(x,Y)\ll 1/\tau$. When
increasing $Y$ in this dilute regime, $n$ rises very fast,
exponentially in $Y$, so like $t$. Around the front position ($x\sim
x_s(Y)$), one has $t\sim \order{1}$ and thus $n\sim\order{1/\tau}$:
on this resolution scale, the target is {\em dense}. Finally, behind
the front, where $t$ saturates at $1$, the occupation number does
{\em not} saturate, rather it keeps growing with $Y$, albeit only
slowly (cf. Eq.~(\ref{eq-dndyhigh})): $n(x,Y)\simeq (Y-Y_c)/\tau$,
where $Y_c\sim \ln(1/\tau)$ is the `critical' rapidity for reaching
saturation in the deposit rate at $x$ : $f(x)=1/\tau$ for $Y>Y_c$.
The precise value of $Y_c$ depends upon $x$ and the initial
conditions at $Y=0$.

\subsection{Particle number fluctuations}

As discussed in Sect. \ref{sect-scatt}, the evolution equations
involve fluctuation terms, like the last term in
Eq.~(\ref{eq-dt2dy}), which reflect the discreteness of the particle
number and are inconsistent with the mean field approximation
underlying Eq.~(\ref{eq-linBK}). Since the occupation numbers are
large at saturation, $n\sim 1/\tau \gg 1$, one may still hope that
the effects of such fluctuations are relatively small and can be
treated in perturbation theory around the mean--field results.
However, this expectation turns out to be naive, as only recently
understood in the context of QCD \cite{IM032,MS04,IMM04,IT04}. The
high--energy evolution is in fact dramatically sensitive to
fluctuations, and the physical reason for that is implicit in the
previous discussion: the evolution is driven by the BFKL growth and
diffusion in the {\em dilute tail} of the front, where the
occupation numbers are small (of order one) and thus the effects of
particle--number fluctuations are important indeed.

The previous analysis of the linear and the mean--field
approximations suggests that the model under consideration falls in
the universality class of the `reaction--diffusion' process
\cite{Saar,Panja} (cf. the Introduction), and this conclusion will
be further supported by the numerical analysis in the next section.
It is then possible --- like in the corresponding studies in QCD
\cite{IMM04,IT04} --- to rely on the correspondence with statistical
physics in order to characterize the {\em universal} aspects of the
dynamics in the presence of fluctuations. These aspects are mostly
{\em qualitative} and refer to the `late--time' (here, large $Y$)
behaviour at `weak coupling' (here, $\tau\ll 1$). In fact, the {\em
semi--quantitative} results which are known to be universal are
valid only for values of $\tau$ which are too small to be relevant
for the analogy with QCD (recall that $\tau = \alpha_s^2$). Yet,
most of the qualitative features that we shall enumerate below will
be later recognized on the numerical results in Sect.
\ref{sect-numerics}.

\vspace*{.2cm}
\begin{enumerate}
\item[\tt(i)] For a given initial condition at $Y=0$, the stochastic
evolution up to $Y$ generates a {\em statistical ensemble of fronts}
(rather than a single front for the deterministic dynamics in the
MFA), with the different fronts in the ensemble differing by their
respective front positions. Hence, $x_s(Y)$ becomes now a {\em
random variable}.

\item[\tt(ii)] To a very good approximation, the distribution of
$x_s$ at $Y$ is a {\rm Gaussian}, with an expectation value
$\avg{x_s}$ and a dispersion $\sigma^2(Y)=\avg{x_s^2}-\avg{x_s}^2$
which rise both linearly with $Y$: $\avg{x_s}=\lambda_s Y$ and
$\sigma^2(Y) = D_{\rm f} Y$. The limitations of this Gaussian
approximation have been analytically studied in Refs.
\cite{BDMM,MSX06} in the limit of high occupancy (or weak coupling),
which amounts to letting $\tau\to 0$ within the toy model.

\item[\tt(iii)] The (asymptotic) value of the {\em
average velocity} $\lambda_s$ is significantly smaller than its
mean--field prediction in Eq.~(\ref{eq-gammas}). The respective
deviation is analytically known in the formal weak--coupling limit
\cite{BD97}, with a result which when adapted to the present toy
model reads
   \beq\label{eq-lambda2}
   \lambda_s \simeq
   \frac{\chi(\gamma_s)}{\gamma_s} -
   \frac{\pi^2 \gamma_s \chi''(\gamma_s)}{2 \ln^2\tau} =
   3 \sqrt{3} - \frac{9 \sqrt{3}\, \pi^2}{2 \ln^2 \tau}\,,
  \eeq
valid when $\ln^2\tau \gg 1$. In the same limit, the {\em front
diffusion coefficient} $D_{\rm f}$ is known to behave like $D_{\rm
f} \sim 1/\ln^3(1/\tau)$ \cite{BDMM}, and thus to vanish, at it
should, when $\tau\to 0$. Notice, however, the very slow, {\em
logarithmic} in $\tau$, convergence of these results to their
respective mean--field limits, which reflects the strong sensitivity
of the evolution to fluctuations.

\item[\tt(iv)] When increasing $Y$, the asymptotic value of $\lambda_s$
is rapidly reached ({\em exponentially} in $Y$), as opposed to the
rather slow convergence predicted by the mean--field result
(\ref{eq-lambda}).

\item[\tt(v)] The individual fronts in the ensemble exhibit
{\em  geometric scaling}, but only over a {\em compact} region: that
is, for each front, there is only a finite distance $x-x_s\le \Delta
x_g$, with $\Delta x_g\simeq (1/\gamma_s)\ln(1/\tau)$, ahead of the
front where the amplitude scales like $t(x)\propto  {\rm
e}^{-\gamma_s (x-x_s)}$, with $\gamma_s= {1}/{\sqrt{3}}$. This
should be contrasted to the mean--field amplitude in
Eq.~(\ref{eq-satsol}), for which the scaling window $x-x_s \propto
\sqrt{Y}$ is ever increasing with $Y$, and hence it can become
arbitrarily large.

\item[\tt(vi)] The width $\Delta x_g$ of the
scaling window in the presence of fluctuations is precisely the
distance over which the amplitude fails from its value $t\simeq 1$
at saturation to a value of $\order{\tau}$, which is the value
corresponding to an occupation number $n(x)$ of $\order{1}$. This is
not an accident: in an actual event, the occupation number per site
is {\em discrete}, $n(x)=0,\,1,\,2,$ etc., so it cannot fall down to
a non--zero value which is smaller than one. Hence, a front in
$n(x)$ has naturally an {\em end point}, which is the rightmost
occupied bin, and where the occupation number is of $\order{1}$.
Now, the corresponding front for $t(x)$ is not truly compact,
because of the non--locality in the elementary scattering amplitude
(recall that $t(x)\approx \int \dif z\, \tau(x|z) n(z)$ in the
dilute regime). However, the latter decays very fast at large
distances, cf. Eq.~(\ref{eq-tau}), so for $x-x_s \gg \Delta x_g$ the
front $t(x)$ develops an exponential tail of the `color
transparency' type: $t(x) \propto  {\rm e}^{-x}$.

\item[\tt(vii)] Physical observables, like the {\em average} dipole
amplitude $\avg{t(x)}_Y$, are obtained by averaging over the
ensemble. In this averaging, the geometric scaling property of the
individual fronts is washed out by the dispersion in the positions
of the fronts \cite{MS04,IMM04}, and is eventually replaced, at
sufficiently large $Y$, by a new type of scaling \cite{IMM04,IT04},
known as {\em diffusive scaling} \cite{HIMST06}. Specifically,
$\avg{t(x)}_Y$ is estimated as
  \beq\label{eq-diffscal}
   \avg{t(x)}_Y \,\simeq\, \frac{1}{2}\,{\rm Erfc} \left[ \frac{x-\avg{x_s}}
   {\sqrt{2D_{\rm f} Y}} \right],
  \eeq
where ${\rm Erfc}(x)$ is the complimentary error function:
 \beq \label{erfcdef}
    {\rm Erfc}(x)\,\equiv\,\frac{2}{\sqrt{\pi}}
    \int\limits_x^\infty {\rm
    d}t\,{\rm e}^{-t^2} \,=\
    \begin{cases}
        \displaystyle{2-\frac{\exp(-x^2)}{\sqrt{\pi}\,|x|}} &
        \text{ for\,  $x \ll -1$}
        \\*[0.1cm]
        \displaystyle{1} &
        \text{ for\,  $|x|\ll 1$}
        \\*[0.1cm]
        \displaystyle{\frac{\exp(-x^2)}{\sqrt{\pi}x}} &
        \text{ for\,  $x \gg 1$}\,.
    \end{cases}
 \eeq
The diffusive scaling approximation, Eq.~(\ref{eq-diffscal}), holds
when $D_{\rm f} Y \gg 1$ and in a wide region around the (average)
saturation line $\avg{x_s}_Y$, such that $|x-\avg{x_s}|\ll
\gamma_sD_{\rm f} Y$. It also smoothly interpolates towards the
black disk limit $\avg{t(x)}= 1$ as $x \to -\infty$. For even larger
distances $x-\avg{x_s}\simge \gamma_sD_{\rm f} Y$, one recovers the
color--transparency behaviour, $\avg{t(x)} \propto \exp
[-(x-\avg{x_s})]$, with a prefactor which is sensitive to
fluctuations (see Refs. \cite{IT04,HIMST06} for details).

\end{enumerate}
\vspace*{.2cm}

\section{Some numerical results}\label{sect-numerics}

As mentioned in the Introduction, our present numerical
investigation of the one--dimensional model is only exploratory,
with the purpose of illustrating some of the analytic results and
qualitative features anticipated in the previous section and,
especially, verifying that the present model lies indeed in the
universality class of the reaction--diffusion process. In this
section, we shall briefly present our numerical technique (a
Monte--Carlo simulation) and then expose the first results obtained
in this way. Further verifications and more detailed results are
deferred to a subsequent publication.

We would like to simulate the particle evolution described by the
master equation (\ref{eq-dpdy}) along with the deposit rate in
Eq.~(\ref{eq-fi}). To that aim, we need to specify the
one--dimensional lattice (that is, the extremal values,
$x_{\text{min}}$ and $x_{\text{max}}$, for $x$, and the lattice
spacing $\Delta$) together with the initial conditions at $Y=0$ and
the value of the parameter $\tau$ which enters the elementary
scattering amplitude \eqref{eq-tau}. In practice, we shall use a
lattice of 1152 size bins between $x_{\text{min}}=-16$ and
$x_{\text{max}}=128$ (hence, $\Delta=0.125$). Also, we choose
$\tau=0.01$, which via the identification $\tau\equiv \alpha_s^2$
corresponds to a value $\alpha_s=0.1$ for the coupling constant in
QCD.

We are merely interested in the universal aspects of the evolution
at large values of $Y$, which we expect to be largely insensitive to
the precise form of the initial conditions. For definiteness, we
choose the particle occupation number at $Y=0$ to be a step
function: $n(x,Y=0)=N_0 \theta(-x)$, with $N_0=2$. For this
distribution, we can compute the initial value of the deposit rate
$f_i$ according to Eqs.~(\ref{eq-fi}) and (\ref{eq-tau}).

To evolve the particle system, we have to simulate particle creation
according to the deposit rate $f_i$. One physical step in this
evolution, which consists in the emission of an additional particle,
involves three steps in the actual algorithm:
\begin{enumerate}
\item we first randomly select the rapidity $\delta Y$ at which the emission
occurs. This is done by generating an exponentially decreasing
distribution in `time' ($Y$) in which the relaxation time is fixed
as the inverse of the total emission rate $\sum_i f_i$;
\item the site $j$ at which the new particle is created is fixed by
randomly selecting a site according to a probability law specified
by the deposit rate (that is, to each site $x_i$, we attribute the
weight function $f_i$);
\item the particle number is increased by one at the selected site $j$
and the deposit rates $f_i$ are updated for all the sites. To speed
up this last step, it is useful to notice that, the only change in
going from $f_i(Y)$ to $f_i(Y+\delta Y)$ refers to the replacement
of $n_j$ by $n_j+1$. This observation together with
Eq.~(\ref{eq-fi}) immediately imply
 \beq
f_i(Y+\delta Y) = \frac{\Delta}{\tau}(1-\sigma_{ij}) + \sigma_{ij}
f_i(Y).
 \eeq
\end{enumerate}
Those steps are repeated up to a maximal rapidity $Y_{\text{max}}$,
that we have chosen as $Y_{\text{max}}=20$. To study the stochastic
aspects of the evolution, we have generated $N_{\text{ev}}=10^5$
distinct events.

\begin{figure}
\subfigure[]{\includegraphics[scale=0.8]{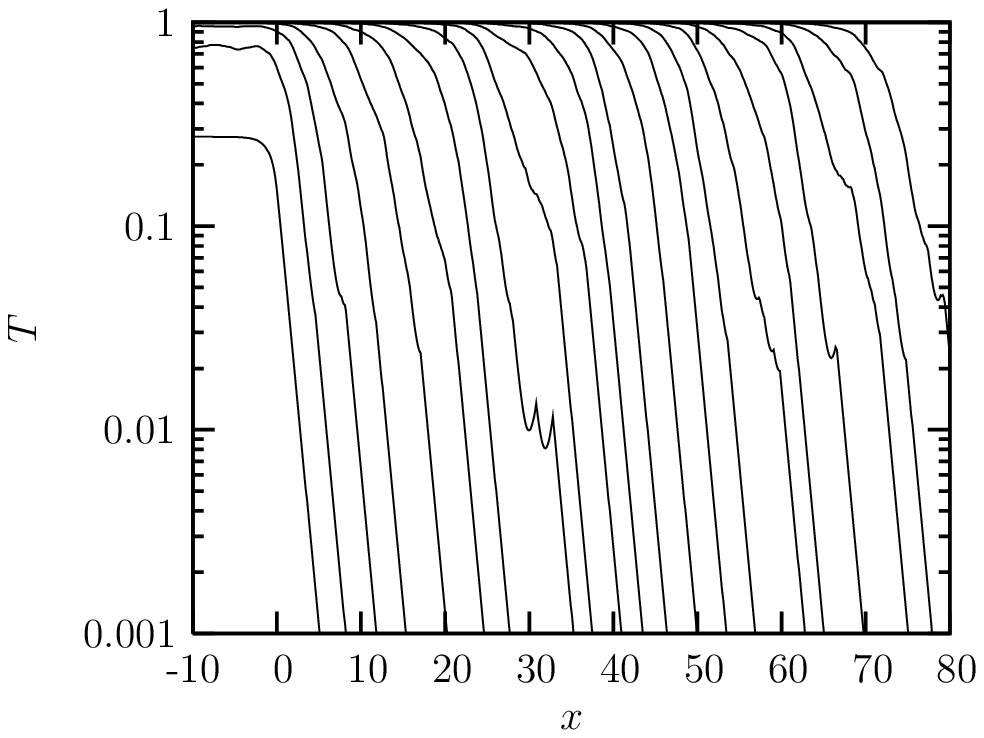}}
\subfigure[]{\includegraphics[scale=0.8]{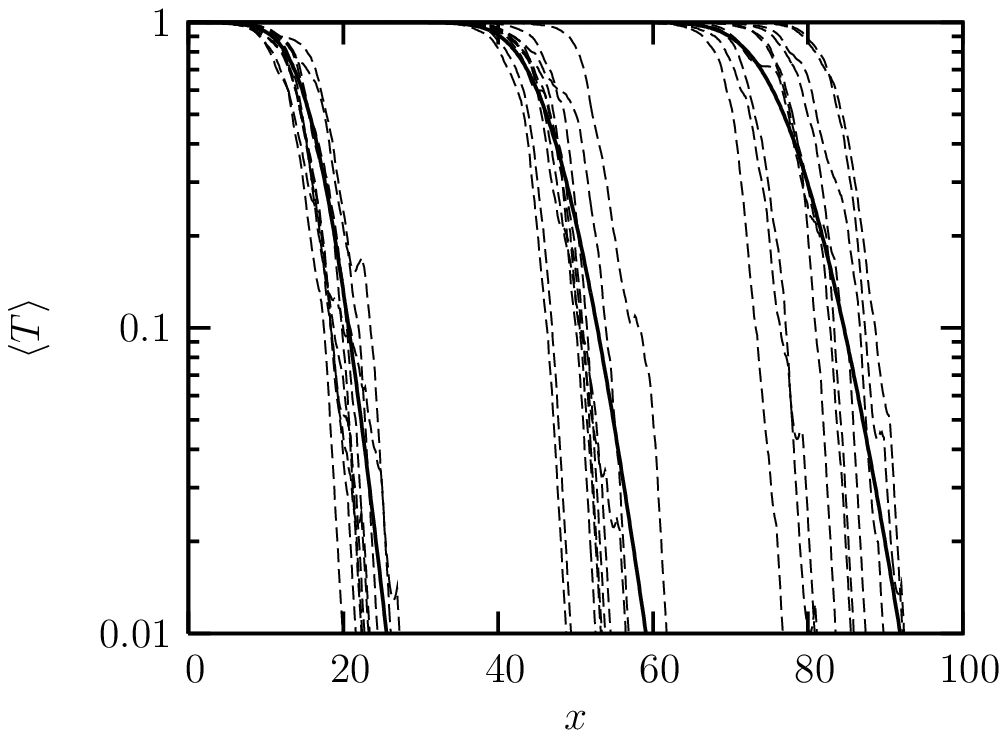}}
\caption{Event-by-event evolution of the scattering amplitude: (a)
Rapidity evolution of the scattering amplitude for a single event.
(b) Amplitude for 10 events (dashed lines) and average amplitude
(solid line) for $Y=5,\,12.5$, and $20$.}\label{fig:event}
\vspace*{.2cm}
\end{figure}

To begin with, let us consider the rapidity evolution of a single
event (see the left plot in Fig. \ref{fig:event}). This demonstrates
the formation of the traveling--wave pattern through two distinct
mechanisms:

\vspace*{.2cm}
\begin{itemize}
\item particles are produced within already occupied sites.
This evolution is basically governed by the mean--field dynamics,
cf. Eq.~(\ref{eq-linBK}), and results in the formation of a
traveling front with the critical parameters discussed in the
previous section; the (average) velocity of this front is however
influenced by the fluctuations to be discussed below (cf.
Eq.~(\ref{eq-lambda2})).
\item from time to time, a rare fluctuation appears far ahead the tip
of the front, in the region that was previously unoccupied. (Such
fluctuations are rare since their probability is exponentially
decreasing with the distance to the already occupied bins, cf.
Eq.~(\ref{eq-tau}).) The subsequent evolution is then a competition
between the local, BFKL--like, growth of that particular fluctuation
and the growth and progression of the mean-field--like wavefront.
When considering a set of events, such fluctuations lead to an
increasing {\em dispersion} in the position of the fronts (see right
plot of Fig. \ref{fig:event}), to which we shall shortly return.
\end{itemize}
\vspace*{.2cm}

\begin{figure}
\subfigure[Number of particle per
bin]{\includegraphics[scale=0.8]{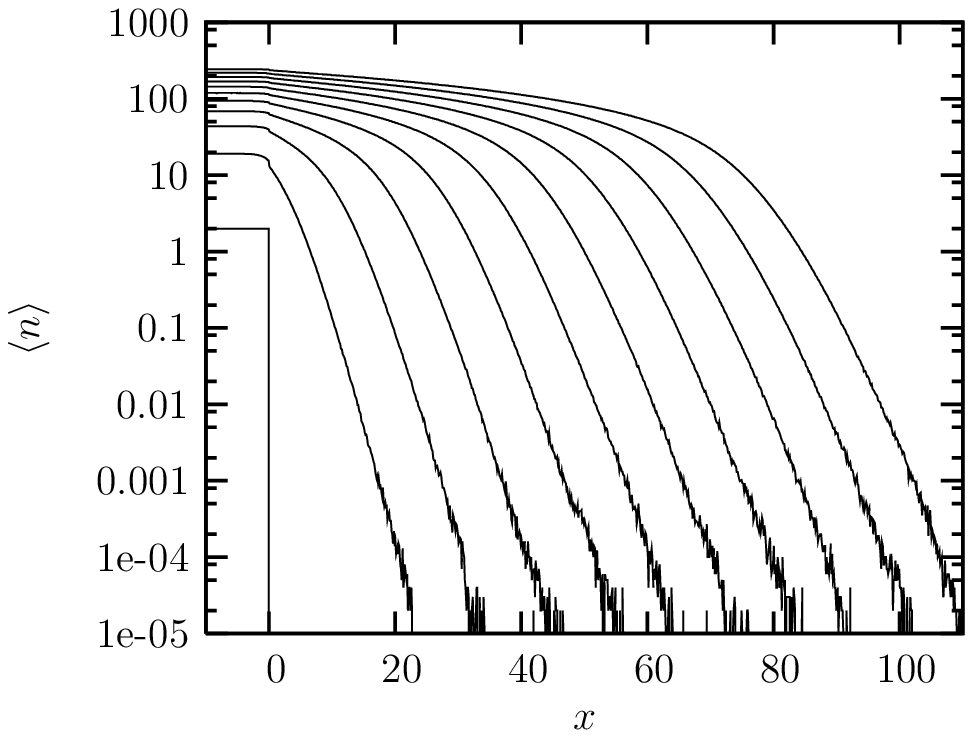}} \subfigure[Scattering
amplitude]{\includegraphics[scale=0.8]{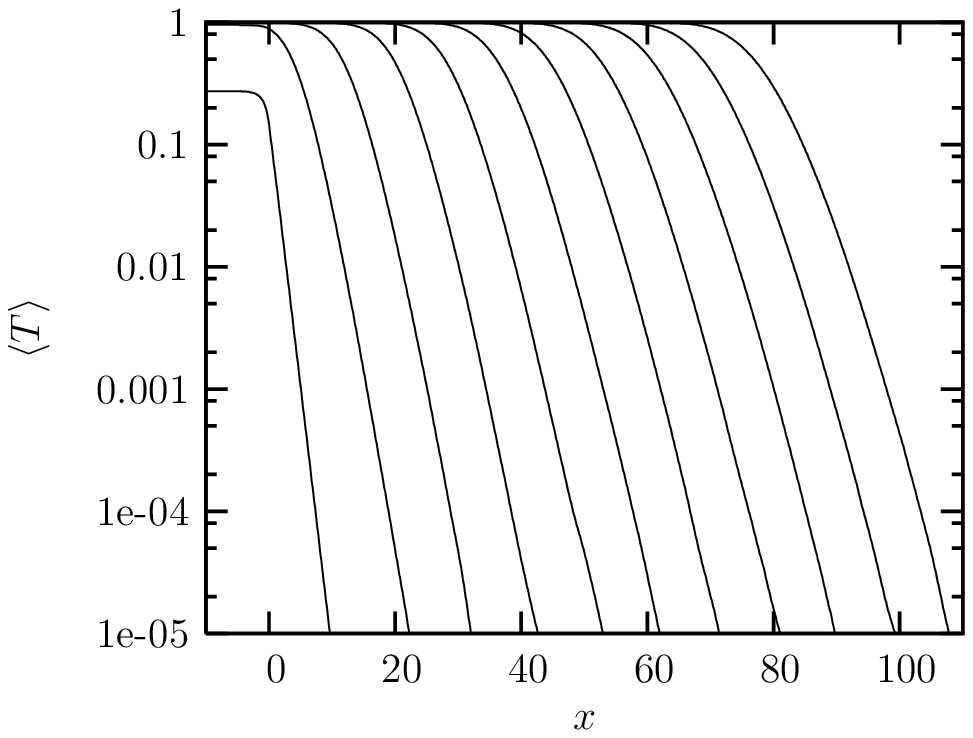}}
\caption{Evolution of the average quantities (particle number and
scattering amplitude) obtained after $10^5$ events. Results are
displayed as a function of $x$ for, from left to right,
$Y=0,2,4,\dots,20$.}\label{fig:avg} \vspace*{.2cm}
\end{figure}

\noindent By averaging over a huge number of events
($N_{\text{ev}}=10^5$), we obtain the averaged distributions shown
in Fig. \ref{fig:avg}, for the average number of particles per
lattice site and for the average scattering amplitude. In the left
plot, we observe large fluctuations in the average number of
particle per site when $\avg{n}\approx 10^{-5} = 1/N_{\text{ev}}$.
Those correspond to rare fluctuations (only a few events have
nonzero occupation number in the respective bins) in the dilute tail
of the front. It is also interesting to notice that the fluctuations
start to be visible around $\avg{n} = 1/\sqrt{N_{\text{ev}}}\approx
0.003$, as expected for particle--number fluctuations. One can also
check that, within the dense regime, the number of particles per bin
increases linearly as expected from Eq.~(\ref{eq-dndyhigh}).

Concerning the average amplitude, the fact that this remains smooth
even in the dilute tail of the front is just because this quantity
is obtained by a convolution of $\avg{n}$ with the elementary
interaction $\tau_{ij}$ (recall that $\avg{t_i}\approx
\sum_j\tau_{ij}\avg{n_j}$ in the dilute regime). Still for the
average amplitude, although one can notice a traveling--wave pattern
in the right plot in Fig. \ref{fig:avg}, a closer look reveals that
the shape of the average front is in fact changing with $Y$ --- its
slope decreases with $Y$. This reflects the violation of geometric
scaling through fluctuations, an important physical effect that we
shall now study in more detail.

\begin{figure}
\subfigure[The average velocity of the traveling
front]{\includegraphics[scale=0.8]{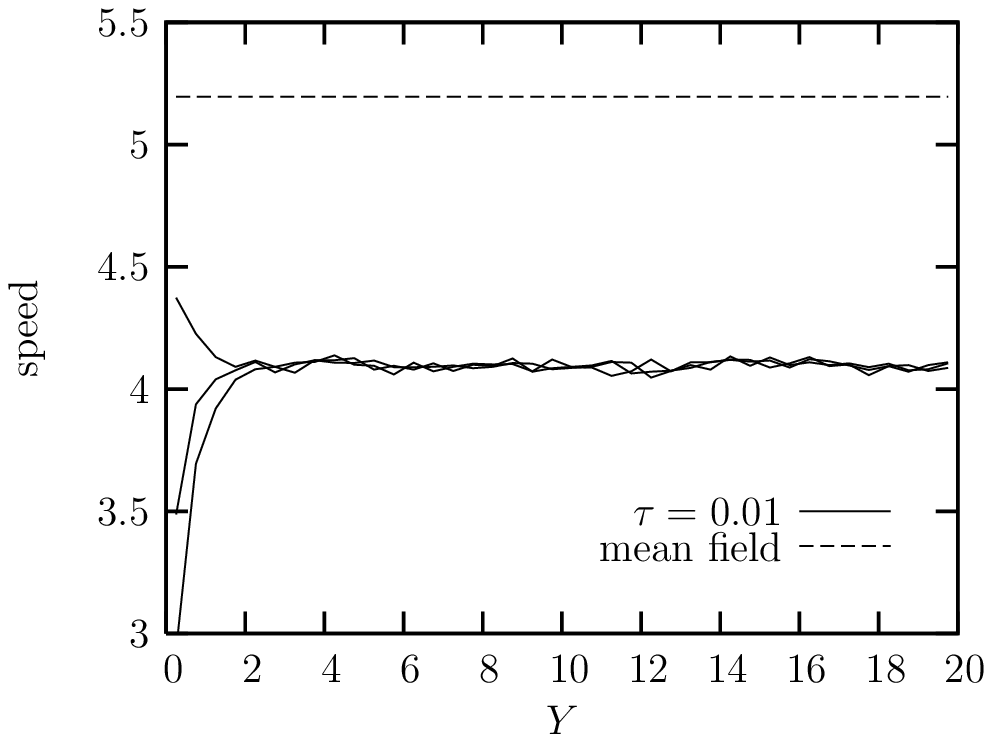}} \subfigure[Dispersion
in the position of the
fronts]{\includegraphics[scale=0.8]{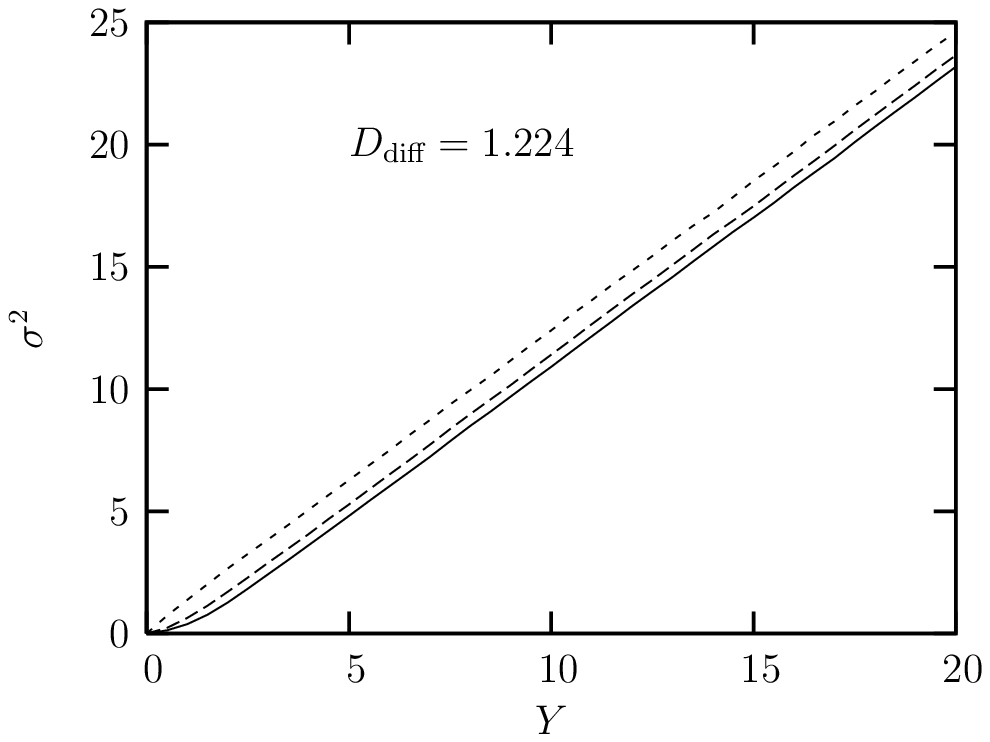}} \caption{The
statistics of the front position (or `saturation scale') $x_s$. Both
the average position and the squared dispersion increase linearly
with rapidity.}\label{fig:stat} \vspace*{.2cm}
\end{figure}

\begin{figure}
\begin{center}
\includegraphics{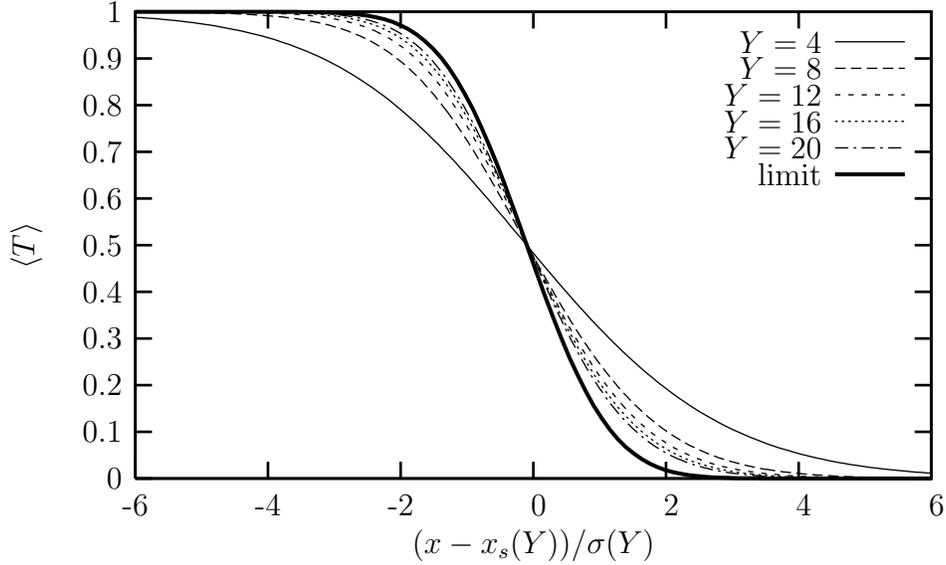}
\end{center}
\caption{The average amplitude represented as a function of the
diffusive scaling variable. When rapidity increases, the asymptotic
result \eqref{eq-diffscal} (indicated in the figure by `limit') is
indeed obtained.}\label{fig:diffscal}\vspace*{.2cm}
\end{figure}

To that aim, we need to first study the statistics of the position
of the front (`the saturation scale $x_s$') as a function of
rapidity. For a given event $t(x,Y)$, the position of the front is
determined by $t(x=x_s(Y),Y) = t_0$, with $t_0$ some constant number
that we have here chosen as $t_0=0.1$ (but we checked that different
choices do not significantly modify the results). From the
discussion in Sect. \ref{sect-analytics}, we expect both the average
value of $x_s(Y)$ and the associated dispersion to grow linearly
with $Y$. For the average value, this is checked on the left plot in
Fig. \ref{fig:stat}, which exhibits our numerical results for
$\partial_Y \avg{x_s(Y)}$. This plot further shows that the
large--$Y$ asymptotic of the velocity is attained pretty fast and,
moreover, this asymptotic value is significantly smaller than the
corresponding prediction of the mean--field approximation, namely $
\lambda_s = 3 \sqrt{3}$ (cf. Eq.~(\ref{eq-gammas})). Next, we turn
the corresponding dispersion, $\sigma^2 = \avg{x_s^2}-\avg{x_s}^2$,
with the results displayed in the right plot in Fig. \ref{fig:stat}.
This is well fitted by a linear increase, $\sigma^2\simeq D_{\rm f}
Y$, with a `front diffusion' coefficient $D_{\rm f}=1.224$. Thus,
the statistics of the `saturation scale' follows indeed the pattern
expected for a `reaction--diffusion' process. We are not in a
position here to also check more quantitative aspects, so like the
slow approach towards the respective mean--field results with
decreasing $\tau$, since the parameter $\tau$ in our calculation not
only has a fixed value, but this value is also too large for
analytic formul\ae{} like Eq.~(\ref{eq-lambda2}) to apply.

We now return to an analysis of the shape of the average amplitude
(cf. the right plot in Fig. \ref{fig:avg}), with the purpose of
verifying another important prediction of the `reaction--diffusion'
problem: the emergence of {\em diffusive scaling} at large $Y$ (cf.
Eq.~(\ref{eq-diffscal})). To check that, we have plotted the average
amplitude obtained from our numerical simulations as a function of
the expected scaling variable $(x-\avg{x_s})/\sigma$, with
$\avg{x_s}$ and $\sigma^2$ taken from our previous analysis (Fig.
\ref{fig:stat}) at various values of $Y$. As manifest on Fig.
\ref{fig:diffscal}, the measured average amplitude converges indeed
towards the expected asymptotic curve, Eq.~(\ref{eq-diffscal}), when
increasing $Y$. The diffusive scaling property is thus fully
satisfied by our model.

\section{Conclusions}
\label{sect-concl} \setcounter{equation}{0}

In this paper we have presented a (1+1)--dimensional stochastic
particle model which mimics high--energy evolution and scattering in
QCD at fixed impact parameter. The `time' coordinate $Y$ in this
model corresponds to rapidity, while the `spatial' coordinate $x$
represents the (logarithm of the) transverse size of a dipole in
QCD. The model is interesting in several respects:

\vspace*{.2cm}
\begin{enumerate}
\item[\tt(i)] The structure of the model is consistent with, and to
a large extend determined by, general physical principles which are
known to hold in QCD: boost--invariance, multiple scattering, and
evolution via the emission of an additional particle per unit
rapidity.

\item[\tt(ii)] Related to the above, the model exhibits a saturation
mechanism which is similar to gluon saturation in QCD: The new
particles created by the evolution are {\em coherently} emitted from
the preexisting ones, with an emission rate which saturates at high
density because of the multiple scattering between the additional
particle and its `parents'.

\item[\tt(iii)] The model appears to be in the universality class of
the reaction--diffusion process, as also expected for QCD, but
unlike other related models in the literature, it does not involve
explicit vertices for particle recombination. These would be
redundant since saturation is anyway ensured by coherence effects in
the particle emission, as explained above.

\item[\tt(iv)] The model exhibits all the qualitative features expected
in QCD at fixed impact parameter, concerning both the mean--field
aspects (a BFKL--like growth in the dilute regime, the formation of
a saturation front with geometric scaling) and the effects of
fluctuations (dispersion in the positions of the fronts, breakdown
of geometric scaling in the statistical description, convergence
towards diffusive scaling at high energy).

\item[\tt(v)] The model is simple enough to allow for detailed
numerical investigations. Hence, it can be used as a playground to
develop and test our intuition concerning the dynamics in QCD at
high energy. Complex phenomena which are expected, but not yet
demonstrated, in QCD (like the emergence of diffusive scaling) can
be visualized in this context and their physical consequences can be
explicitly analyzed.

\item[\tt(vi)] The structural aspects of the model are particularly
interesting since they may inspire our searches for corresponding
structures in QCD. In particular, the evolution equations for the
scattering amplitudes in this model appear as a natural
generalization (within the limits of the model, of course) of the
Balitsky--JIMWLK equations in which the projectile and target are
{\em symmetrically} treated: for each of these two systems, the
equations include particle--number fluctuations and multiple
scattering with both the particles internal to the system
(saturation effects) and those in the other system (unitarity
corrections).

\end{enumerate}
\vspace*{.2cm}

In the remaining part of this concluding section we would like to
elaborate a little bit more on the last point above and
comparatively discuss the evolution equations in the toy model and
the `Pomeron loop' equations proposed in QCD at large $N_c$
\cite{IT04,MSW05,IT05}. This comparison is discussed in more detail
in the Appendix from which we shall extract here the main
conclusions.

By inspection of the two sets of equations --- see, e.g.,
Eqs.~(\ref{eq-dtxdy}), (\ref{eq-dt2dy}) and (\ref{eq-dt3dy}) for the
toy model, and, respectively,  Eqs.~(2.7)--(2.8) in Ref. \cite{IT05}
for QCD ---, one can notice some important differences which
formally refer to the structure and physical interpretation of the
respective fluctuation terms: Within the toy model, we have seen
(cf. Eq.~(\ref{eq-fluct})) that these terms correspond to the
multiple scattering of {\em only one} among the two child particles
produced by a splitting in the target with two (or more) particles
from the projectile. By contrast, these particular
multiple--scattering effects are not at all included in the
respective `Pomeron loop' equations, where the fluctuation terms
rather refer to the {\em separate} scattering of {\em both} child
dipoles produced after a splitting in the target with two
(different) dipoles in the projectile. However, as explained in the
Appendix, such differences seem to be inessential (except in the
very early stages of the evolution) since they amount to a
reorganization of the perturbation theory in the dilute regime,
which however leads to the same dominant behaviour at sufficiently
high energy --- namely, such that $Y\simge 1$ in the toy model and,
respectively, $\abar Y\simge 1$ in QCD.

To briefly explain this reorganization, let us consider the
weak--scattering regime, where the scattering amplitude for a
projectile made with $k$ particles can be estimated as (for $Y\simge
1$)
   \beq\label{eq-tkY}
 \avg{t^{(k)}(x_1,\dots ,x_k)}_Y\,=\,
  \int\limits_{z_i} \tau(x_1|z_1) \cdots
  \tau(x_k|z_k)\avg{n^{(k)}(z_1,\dots ,z_k)}_Y,\eeq
where $\lan{n^{(k)}}\ran$ is the `normal--ordered' $k$--body density
of Eq.~(\ref{eq-nk}). This formula assumes that the $k$ external
particles scatter off $k$ {\em different} particles in the target,
and thus treats the target and the projectile on the same footing
--- multiple scattering is neglected for both of them. Strictly
speaking, Eq.~(\ref{eq-tkY}) is not the same as the $k$--particle
scattering amplitude in fixed order perturbation theory (the latter
would include additional contributions associated with the multiple
scattering of individual target particles; see, e.g.,
Eq.~(\ref{eq-tt}) in the Appendix), but it becomes equivalent to the
latter when $Y\simge 1$ since it captures the contribution which has
the fastest rise with $Y$ (the $k$--Pomerons piece of the complete
amplitude).

In QCD at low density (and large $N_c$), one can define `bare'
amplitudes similar to Eq.~(\ref{eq-tkY}) in terms of dipoles and
then deduce linear evolution equations for these quantities within
the dipole picture\cite{IT05}. The `Pomeron loop' equations are then
obtained by completing these linear equations with the non--linear
terms responsible for unitarity corrections, as taken over from the
Balitsky--JIMWLK equations. This procedure amounts to dressing the
`bare' amplitudes $\lan{t^{(k)}}\ran$ with multiple--scattering
effects {\em on the side of the projectile}. In the toy model, on
the other hand, the evolution equations obtained in Sect.
\ref{sect-scatt} correspond to dressing $\lan{t^{(k)}}\ran$ with
multiple scattering {\em on both the projectile, and the target,
sides}.

Clearly, the last procedure is conceptually more satisfactory, as it
provides a symmetric treatment of the target and the projectile in
the general, strong--scattering, regime. In QCD, such a symmetric
description is still lacking and its construction remains as an
important open problem. But although conceptually, and also
aesthetically, more appealing, it is not clear to us whether such a
symmetric description would differ indeed
--- in terms of {\em physical consequences} at high energy ---
from that provided by the already known `Pomeron loop' equations.

\section*{Acknowledgments}
G.S. is funded by the National Funds for Scientific Research (FNRS,
Belgium). J.T.S.A. is partially funded by CNPq and CAPES, Brazil.

\appendix

\section{On the fluctuation terms in the Pomeron loop equations}
\label{sect-app} \setcounter{equation}{0}

In this Appendix we shall describe and clarify some structural
differences between the evolution equations generated by the toy
model and the Pomeron loop equations in QCD \cite{IT04,MSW05,IT05}.
To that aim, it is useful to also have under ones eyes the third
equation in the toy--model hierarchy, as obeyed by $\avg{t_x
t_yt_z}$; this reads:
  \beq\label{eq-dt3dy}
   \frac{\del \avg{t_x t_y t_z}}{\del Y} = &&
   \int\limits_{w} \frac{\tau_{xw}}{\tau}
   \avg{t_w t_y t_z (1- t_x)}\, +\, \text{permutations}
   \nn + &&
   \int\limits_{w} \frac{\tau_{xw}\tau_{wy}}{\tau}
   \avg{t_w t_z (1- t_x)(1-t_y)}\, +\, \text{permutations}
   \nn + &&
   \int\limits_{w} \frac{\tau_{xw}\tau_{yw}\tau_{zw}}{\tau}
   \avg{t_w (1- t_x)(1-t_y)(1-t_z)}.
  \eeq

Already a superficial comparison between the toy--model equations,
cf. Eqs.~(\ref{eq-dtxdy})--(\ref{eq-dt2dy}) and (\ref{eq-dt3dy}),
and the corresponding equations in QCD, cf. Ref. \cite{IT05},
reveals some important dissimilarities that we summarize here:

\vspace*{.2cm} \texttt{(A)} The equation for $\lan t^k\ran\equiv
\avg{t_{x_1}\dots t_{x_k}}$ within the toy model involve several
types of fluctuation, with the following generic structures: $\tau
\lan t^{k-1} \ran$, $\tau^2 \lan t^{k-2} \ran$, $\dots$, $\tau^{k-1}
\lan t \ran$. These terms are all of the same order in the low
density regime where $\avg{t} \sim \tau$, and hence contribute on
equal footing to building up the many--body correlations. By
contrast, the corresponding\footnote{Note that we use the
upper--case notation $T$ for the scattering amplitude of a dipole in
QCD to distinguish from the respective amplitudes $t$ in the toy
model.} `Pomeron loop' equation for $\lan T^{(k)}\ran$ involves just
one fluctuation term, with the generic structure $\tau\lan
T^{(k-1)}\ran$ (where $\tau\sim\order{\alpha_s^2}$ in QCD).

\texttt{(B)} As discussed in relation with Eq.~(\ref{eq-fluct}), the
fluctuation term in the equation for $\avg{t_x t_y}$ corresponds to
the double scattering of one of the daughter particles produced by a
splitting in the target. A similar conclusion applies to the higher
equations in the toy--model hierarchy, like Eq.~(\ref{eq-dt3dy}): a
term like $({\tau_{xw}\tau_{wy}}/{\tau})\avg{t_w t_z}$ in the second
line of Eq.~(\ref{eq-dt3dy}) describes a process in which the two
external particles at $x$ and $y$ scatter off the daughter particle
produced at $w$ after a splitting in the target (with the third
external particles at $z$ being just a spectator); also, a term like
$({\tau_{xw}\tau_{yw}\tau_{zw}}/{\tau})\avg{t_w}$ in the third line
there corresponds to a {\em triple} scattering for the target
particle at $w$, which now scatters off all the external particles.

By contrast, in the construction \cite{IT04,MSW05,IT05} of the
fluctuation terms in the Pomeron loop equations in QCD, one has
completely neglected the multiple scattering for the individual
target dipoles. Rather, the fluctuation terms included in these
equations describe the separate scattering of {\em both} child
dipoles produced after a splitting in the target with a pair of
projectile dipoles.

\vspace*{.2cm}

The above discussion rises several questions:

\vspace*{.2cm}
\begin{enumerate}
\item[\tt(i)] Why there are no fluctuation terms in the toy--model
equations corresponding to the {\em single} scattering of both
particles $w$ and $z$ produced after a splitting $w\to wz$ in the
target ?

\item[\tt(ii)] Was it correct to neglect multiple scattering for the
individual target dipoles in the corresponding equations in QCD ?

\item[\tt(iii)] What should be the complete structure of the fluctuation
terms in  QCD ?
\end{enumerate}
\vspace*{.2cm}

We are not so ambitious to try and fully answer question $\tt(iii)$
in what follows, but we would like to clarify at least the answers
to the first two questions. Specifically, we will show that the
reason why the two sets of equations look so different is because
they apply to {\em different} quantities which however have the {\em
same dominant behaviour} at high energy (namely, for $Y\simge 1$).

The following considerations will be restricted to a {\em dilute
target}, off which the individual projectile particles scatter only
once. It is then preferable to express the scattering amplitudes in
terms of particle densities in the target and follow the evolution
of the latter. By using
 \beq t(x)\equiv 1- \exp \left[
   \int \dif z\, n(z) \ln \sigma(x|z) \right]
   \approx \int\dif z\, \tau(x|z)\,n(z)
   \eeq
(with the last, approximate, equality valid in the dilute regime),
together with the definition (\ref{eq-n2}) of the normal--ordered
pair density, one can immediately deduce
 \beq\label{eq-tt}
 \avg{t_x t_y} \approx \int\limits_{z,w} \tau_{xz} \tau_{yw}\,\avg{n_zn_w}
  \,=\,\int\limits_{z,w} \tau_{xz} \tau_{yw}\avg{n^{(2)}_{zw}} +
  \int\limits_{z} \tau_{xz} \tau_{yz}\avg{n_z},\eeq
where the contribution proportional to $\langle n^{(2)}\rangle $
describes the scattering between the two projectile particles and
two {\em different} target particles, while that involving $\avg{n}$
describes the double scattering of a {\em same} target particle. We
shall simultaneously consider the following quantity
  \beq\label{eq-ttY}
 \avg{t^{(2)}_{xy}}\,=\,
  \int\limits_{z,w} \tau_{xz} \tau_{yw}\avg{n^{(2)}_{zw}},\eeq
which represents the toy--model analog of the 2--dipole scattering
amplitude $\lan T^{(2)}\ran$ which enters the `Pomeron loop'
equations in QCD \cite{IT04,IT05}. As compared to Eq.~(\ref{eq-tt}),
the double--scattering terms are neglected in Eq.~(\ref{eq-ttY}),
which therefore treats more symmetrically the target and the
projectile in this weak--scattering regime (in the sense that all
the particles undergo single scattering). However, it is a priori
the less symmetric formula, Eq.~(\ref{eq-tt}), which yields the
2--particle scattering amplitude at fixed order in perturbation
theory: indeed, both terms in the r.h.s. of Eq.~(\ref{eq-tt}) are of
$\order{\tau^2}$. A similar discussion applies to the $k$--particle
amplitudes --- $\lan t^k\ran$ and respectively $\langle
t^{(k)}\rangle$ --- with $k\ge 3$.

In what follow we shall use the evolution equations obeyed by the
particle densities in the dilute regime to deduce the respective
equations for $\lan t^2\ran$ and $\langle t^{(2)}\rangle$. For the
former, we shall recover, as expected, the linear part of the
general equation (\ref{eq-dt2dy}), but with a more transparent
interpretation for the fluctuation terms, which will also answer
question $\tt(i)$ above. For the latter, we shall find the
toy--model analog of the second equation in the `Pomeron loop'
hierarchy.

The low--density versions of Eqs.~(\ref{eq-dndycont}) and
(\ref{eq-dndy2cont}) read
 \beq
  \label{eq-dndyfl}
  \frac{\del \avg{n_{z}}}{\del Y} &\approx&
 \int\limits_u \frac{\tau_{zu}}{\tau}\,\avg{n_u}\,,\\
 \label{eq-dndy2fl}
 \frac{\del \avg{n^{(2)}_{zw}}}{\del Y} &\approx&
 \int\limits_u \left[\frac{\tau_{zu}}{\tau}\, \avg{n^{(2)}_{uw}}\,
 +\, \frac{\tau_{uw}}{\tau}\, \avg{n^{(2)}_{zu}}\right] \,+\,
 \frac{\tau_{zw}}{\tau}\big(\avg{n_z}+\avg{n_w}\big).
  \eeq
Notice the terms linear in $\avg{n}$ in the r.h.s. of
Eq.~(\ref{eq-dndy2fl}): these are {\em fluctuation terms} which
generate a pair $(z,w)$ of particles in the target via the splitting
of a single original particle at $z$, or at $w$. These terms are
very similar to those generated by the dipole picture in QCD
\cite{IT04}.

The equation obeyed by $\avg{t_x t_y}$ can now be obtained by taking
a derivative w.r.t. $Y$ in Eq.~(\ref{eq-tt}) and then using
Eqs.~(\ref{eq-dndyfl})--(\ref{eq-dndy2fl}). By inspection of these
equations, one may expect this procedure to produce {\em two} types
of `fluctuation terms' (i.e., terms linear in $\avg{t}$) : those
generated by the BFKL evolution of the average particle density, cf.
Eq.~(\ref{eq-dndyfl}), and the `genuine' fluctuation terms coming
from the evolution of the pair density, cf. Eq.~(\ref{eq-dndy2fl}).
The latter would correspond to the separate scattering of {\em both}
particles $w$ and $z$ produced by the splitting in the target, and
thus would be the toy--model analog of the fluctuation terms which
appear in the Pomeron loops equations in QCD. However, from the
previous discussion of Eq.~(\ref{eq-fluct}), we know already that
the relevant fluctuation term comes {\em fully} from the BFKL
evolution of $\avg{n}$,  cf. Eq.~(\ref{eq-dndyfl}) ! What happens
then to the fluctuation terms which are explicit in
Eq.~(\ref{eq-dndy2fl}) ? As we explain now, they are in fact
absorbed --- via Eq.~(\ref{eq-tt}) --- in the structure of the
2--particle amplitudes which enter the `BFKL' terms in
Eq.~(\ref{eq-dt2dy}). Indeed, Eqs.~(\ref{eq-tt}) and
(\ref{eq-dndy2fl}) can be used to deduce
 \beq
 \int\limits_{z,w} \tau_{xz} \tau_{yw}\,
 \frac{\del \avg{n^{(2)}_{zw}}}{\del Y}&=&
 \int\limits_{u,z,w}  \tau_{xz} \tau_{yw}
 \left[\frac{\tau_{zu}}{\tau}\,\avg{n_un_w}
 \,
 +\, \frac{\tau_{uw}}{\tau}\,\avg{n_zn_u}\right]\nn
 &=&
 \int\limits_z \left[
   \frac{\tau_{xz}}{\tau} \avg{t_zt_y}+
   \frac{\tau_{yz}}{\tau} \avg{t_zt_x}
   \right]\!,
 \eeq
where in writing the first equality we have already reabsorbed the
fluctuation terms into 2--body densities like $\avg{n_u n_w}$, via
Eq.~(\ref{eq-n2}). In the second line of the above equation one can
recognize the BFKL terms from Eq.~(\ref{eq-dt2dy}), as anticipated.

This answers question $\tt(i)$ above --- the analogs of the
fluctuation terms in the `Pomeron loop' equations are {\em
implicitly} included in the toy--model equations, as
multiple--scattering contributions to the amplitudes which enter the
BFKL terms there --- but rises new questions in return : Why are
these terms explicit in the QCD equations in Ref.
\cite{IT04,MSW05,IT05}, while they are implicit in the corresponding
toy--model equations ? And why the fluctuations terms which are
explicit in the toy model (those corresponding to multiple
scattering off a same target particle) appear to be absent in QCD ?

The answer to these last questions, and also to the original
question $\tt(ii)$, is that the `Pomeron loop' equations are in fact
written for {\em different} quantities, namely the `reduced'
scattering amplitudes $\langle t^{(k)}\rangle$ (see
Eq.~(\ref{eq-ttY})) which in this dilute regime neglect multiple
scattering for both the target particles and the projectile ones.
This amounts to a different reorganization of the perturbation
theory in the weak--scattering regime, which however leads to the
same high--energy behaviour, as previously argued in Refs.
\cite{MMSW05,HIMS05}. (The subsequent discussion is in fact adapted
from Ref. \cite{HIMS05}.) To see this, let us consider the relative
importance of the two terms in Eq.~(\ref{eq-tt}). For more clarity,
assume that the target starts at $Y=0$ as a single particle located
at $x_0$. Hence $\avg{n_{z}}_0=\delta(z-x_0)$ and
$\avg{n^{(2)}_{zw}}_0=0$. Then, clearly, the term  proportional to
$\avg{n}$ in Eq.~(\ref{eq-tt}) will dominate in the early stages of
the evolution ($Y\ll 1$), until a non--vanishing pair density gets
first created via fluctuations, cf. Eq.~(\ref{eq-dndy2fl}). But once
the latter becomes non--zero, its subsequent evolution is much
faster
--- it rises as a `double--Pomeron', $\lan{n^{(2)}}\ran\sim
\exp(2\omega_{\mathbb{P}} Y)$ with $\omega_{\mathbb{P}} = 2$, while
the single--density rises only as $\lan{n}\ran\sim
\exp(\omega_{\mathbb{P}} Y)$ --- and thus it dominates the
2--particle amplitude in Eq.~(\ref{eq-tt}) for all rapidities
$Y\simge 1$.

To summarize, in the high--energy regime where $Y$ is relatively
large, $Y\simge 1$, but not {\em too} large (so that we are still in
the {\em weak scattering regime}; this condition requires $Y\ll
\ln(1/\tau)$), the 2--particle scattering amplitude can be estimated
simply as $\lan{t^{(2)}}\ran$, cf. Eq.~(\ref{eq-ttY}). Equivalently,
this can be obtained by solving the evolution equation for
$\lan{t^{(2)}}\ran$ (see below) with the initial condition
$\lan{t^{(2)}}\ran_0=0$. Note that this initial condition is {\em
different} from that to be used in relation with the original
equation (\ref{eq-dt2dy}), which follows from Eq.~(\ref{eq-tt}) and
reads $\avg{t_{x}t_y}_0=\tau(x|x_0)\tau(y|x_0)$.

As quite clear by inspection of Eqs.~(\ref{eq-ttY}) and
(\ref{eq-dndy2fl}), the fluctuation terms in the equation for
$\avg{t^{(2)}_{xy}}$ come fully from the process in which the two
external particles scatter off {\em both} child particles produced
by a splitting in the target. Specifically, one finds
   \beq\label{eq-dttdY}
  \frac{\del \avg{t^{(2)}_{xy}}}{\del Y}\,=\,
   \int\limits_z \,\left[
   \frac{\tau_{xz}}{\tau} \avg{t^{(2)}_{zy}}+
   \frac{\tau_{yz}}{\tau} \avg{t^{(2)}_{xz}}\right]\,
  + \,\frac{\del \avg{t^{(2)}_{xy}}}{\del
   Y}\bigg |_{\rm fluct}
    \eeq
with the following expression for the fluctuation term:
 \beq\label{eq-dttdYfluct}
 \frac{\del \avg{t^{(2)}_{xy}}}{\del
   Y}\bigg |_{\rm fluct}&=&  \int\limits_{z,w}  \big(\tau_{xz} \tau_{yw}
  +\tau_{xw} \tau_{yz}\big)\,\frac{\tau_{wz}}{\tau}\avg{n_z}\nn
  &=&  \int\limits_{z,w,u}  \big(\tau_{xz} \tau_{yw}
  +\tau_{xw} \tau_{yz}\big)\,\frac{\tau_{wz}}{\tau}\,\tau^{-1}_{zu}
  \avg{t_u}\,,
    \eeq
where in the second line we have expressed the average particle
density in the target $\avg{n_z}$ in terms of the one--particle
scattering amplitude $\avg{t_u}$ by inverting the relation
$\avg{t_u}=\int_z \tau_{uz}\avg{n_z}$, valid in the dilute regime.
The inverse $\tau^{-1}_{xy}$ of the elementary scattering amplitude
exists indeed, since $\tau_{xy}$ is positive--definite; in
particular, for the function $\tau_{xy}$ in Eq.~(\ref{eq-tau}), one
has
 \beq
 \tau^{-1}(x|y)\,=\,\frac{1}{2\tau}\big(1-\del_x^2\big)\delta(x-y).
 \eeq
As anticipated, Eq.~(\ref{eq-dttdYfluct}) is the toy--model analog
of the fluctuation term in the corresponding `Pomeron loop' equation
(see Eq.~(2.7) in Ref. \cite{IT05}). This analogy extends, of
course, to the higher equations in the respective hierarchies.

We have previously noticed that, for $Y\simge 1$, there is no
important loss of accuracy if the solution $\avg{t^{(2)}_{xy}}$ to
Eqs.~(\ref{eq-dttdY})--(\ref{eq-dttdYfluct}) with initial condition
$\lan{t^{(2)}}\ran_0=0$ is used instead of the solution
$\avg{t_{x}t_y}$ to the more general equation (\ref{eq-dt2dy}) with
initial condition $\avg{t_{x}t_y}_0=\tau(x|x_0)\tau(y|x_0)$. By the
same token we deduce that, within QCD, the solutions to the Pomeron
loop equations with initial conditions $\lan{T^{(k)}}\ran_0=0$ for
$k\ge 2$ should provide the correct physical behaviour at high
energy. This conclusion is further supported by the arguments in
Ref. \cite{MMSW05,HIMS05}.



\end{document}